\documentclass[useAMS,usegraphicx,usenatbib]{mn2e}
\usepackage{color}
\usepackage[space]{grffile}

\def\lesssim{\,\lower2truept\hbox{${<\atop\hbox{\raise4truept\hbox{$\sim$}}}$}\,
}
\def\gtrsim{\,\lower2truept\hbox{${>\atop\hbox{\raise4truept\hbox{$\sim$}}}$}\,}

\def\rev{}

\newcommand{\hmass}{~h^{-1}~{\rm M}_\odot}
\newcommand{\msunyr}{$\rm{M}_\odot \, \mbox{yr}^{-1}$}

\title[The early phases of galaxy clusters formation in IR]{The early phases of galaxy
clusters formation in IR: coupling hydrodynamical simulations with GRASIL-3D}

\author[Granato et al.]{
\parbox[t]{\textwidth}{
Gian~Luigi Granato$^{1}$\thanks{Email: granato@oats.inaf.it},
Cinthia Ragone-Figueroa$^{2,1}$\thanks{Email: cin@oac.unc.edu.ar},
Rosa Dom\'{\i}nguez-Tenreiro$^{3}$,
Aura Obreja$^{3}$,
Stefano Borgani$^{1,4}$,
Gabriella De Lucia$^{1}$ and
Giuseppe Murante$^{1}$
}
\vspace*{6pt} \\
  $^1$ Istituto Nazionale di Astrofisica INAF, Osservatorio Astronomico di
  Trieste, Via
  Tiepolo
  11, I-34131
  Trieste, Italy \\
  $^2$ Instituto de Astronom\'ia Te\'orica y Experimental (IATE),\\
  Consejo Nacional de Investigaciones Cient\'ificas y T\'ecnicas de la
Rep\'ublica Argentina (CONICET),\\
  Observatorio
  Astron\'omico, Universidad Nacional de C\'ordoba, Laprida 854, X5000BGR,
C\'ordoba, Argentina\\
$^{3}$Depto. de F\'{i}sica Te\'orica, Universidad Aut\'onoma de Madrid, E-28049 Cantoblanco Madrid, Spain\\
  $^4$ Astronomy Unit, Department of Physics, University of Trieste, via Tiepolo 11, I-34131 Trieste, Italy\\
}

\begin{document}

\newcommand{\gtd}{GRASIL-3D}
\newcommand{\cinthia}[1]{{\bf\textcolor{red}{ #1}}}
\date{Accepted 2015 March 25.  Received 2015 March 17; in original form 2014 December 18}

\maketitle

\begin{abstract}
We compute and study the infrared and sub-mm properties of high redshift ($z
\gtrsim 1$) simulated clusters and proto-clusters. The results of a large set
of hydro-dynamical zoom-in simulations including active galactic nuclei (AGN)
feedback, have been treated with the recently developed radiative
transfer code GRASIL-3D, which accounts for the effect of dust reprocessing
in an arbitrary geometry. {\rev Here, we have slightly generalized the code
to adapt it to the present purpose. Then we have post-processed boxes of
physical size 2 Mpc encompassing each of the 24 most massive clusters
identified at z=0, at several redshifts between 0.5 and 3, producing IR and
sub-mm mock images of these regions and SEDs of the radiation coming out from
them.}

While this field is in its infancy from the observational point of view,
rapid development is expected in the near future thanks to observations
performed in the far IR and sub-mm bands. Notably, we find that in this
spectral regime our prediction are little affected by the assumption required
by this post-processing, and the emission is mostly powered by star formation
rather than accretion onto super massive black hole (SMBH).

The comparison with the little observational information currently available,
highlights that the simulated cluster regions never attain the impressive
star formation rates suggested by these observations. This problem becomes
more intriguing taking into account that the brightest cluster galaxies
(BCGs) in the same simulations turn out to be too massive.
It seems that the interplay between the feedback schemes and the
star formation model should be revised, possibly incorporating a positive
feedback mode.

\end{abstract}


\section{Introduction}
\label{sec:intro}

%Cluster of galaxies, the most massive collapsed structures in the Universe,
%comprise up to several thousands of galaxies, embedded in an huge dark matter
%halo, which is permeated by an hot Intra Cluster Medium. The latter emits in
%the X-rays and up-scatters the CMB photons, and accounts for the bulk of their
%baryonic content.

Galaxy clusters are fundamental probes for many questions of extra-galactic
astrophysics and cosmology, both from the observational as well as from the
theoretical point of view \citep[for a recent review
see][]{kravtsov_borgani12}. Their properties are relatively well known at low
redshift $\lesssim 1$. In particular, it has been assessed that the central
regions of massive clusters are dominated by passive early type galaxies.
Their stellar populations are old, having formed at $z \gtrsim 2$. More
debated and uncertain is the main epoch at which these stellar populations
assembled into the single galactic units we see in the local universe.
Indeed, there seems to be some disagreement between the relatively important
role of merging below $z\lesssim 1$ expected on the basis of recent
theoretical computations \citep{guo11,johansson12}, and some observational
constraints \citep{stott11,lidman12,lin13,inagaki14}. Unsurprisingly,  a
proper theoretical understanding of this and other questions concerning the
evolution of galaxy populations in clusters requires a treatment of the
various processes driving it, such as star formation, AGN activity, feedback
and dynamical interactions at a much deeper level than presently feasible.
Galaxy clusters over cosmic time are prime laboratories for these processes
acting together, which results in a clear environmental dependence of the basic
properties of galaxies, and of their evolutionary history.

At $z \gtrsim 1$, a regime approaching the formation epoch of
massive clusters, observational studies become more and more
problematic and, as a consequence, scarce.  The identification of clusters
(or proto-clusters, somewhat loosely defined as systems that exhibit a
signiﬁcant over-density of galaxies, not yet  gravitationally bound, but that
may collapse to form a cluster at later time) becomes difficult, due both to
increasing detection challenges and intrinsic rareness. Also, a detailed
determination of the properties of their galaxies is progressively uncertain.

Nevertheless, the study of high z clusters is now an active field of
research, as testified by the fact that even the redshift barrier of z=1.5
has been broken in the last few years, by means of mid infrared or X-ray
selection, albeit by just a handful of examples so far. These observations
suggest that, while  galaxy populations  in  the  centers  of  massive
clusters show little change out to $z \sim 1.5$ \citep{mei09,strazzullo10},
in higher redshifts clusters, intense star formation becomes common,
even in the cores and in the most massive galaxies
\citep{tran10,hilton10,
hayashi11,santos11,fassbender14,dannerbauer14,santos14,santos14b}. For
instance, \cite{tran10}, considering only the IR luminous galaxies in the
core (projected distances $< 0.5 {\rm Mpc}$) of a  {\it Spitzer}-selected
cluster at $z=1.62$, found that the star formation rate (SFR) {\rev surface}
density to be at least $1700~ \mbox{M}_\odot~ {\rm yr}^{-1}~ {\rm Mpc}^{-2}$.
This estimate has been later revised downward by \cite{santos14} to $990~ \pm
120 \ \mbox{M}_\odot~ {\rm yr}^{-1}~ {\rm Mpc}^{-2}$, with a contribution
from the BCG of $256~ \pm 70$ \msunyr. \cite{santos14b} measured a strikingly
high amount of star formation $\sim$ 1100 \msunyr\ in the inner 250 kpc  of a
massive ($\sim 5 \times 10^{14} \mbox{M}_\odot$) cluster at $z\sim 1.6$
\cite{dannerbauer14}, using APEX LABOCA 870 $\mu$m observations of the field
around the so-called spiderweb radio galaxy at z=2.16, widely studied as a
signpost for a massive cluster in formation, measured a SFR density of $\sim$
900 \msunyr Mpc$^{-3}$, occurring within a region of 2 Mpc
\footnote{\label{note:imf} The SFRs reported in this paragraph have been
obtained assuming a relationship between the total IR luminosity and SFR. The
calibration of this relationship depends on the assumed initial mass function
(IMF), which in the original papers was Salpeter in some cases
\citep{dannerbauer14,santos14b} and Chabrier  in others
\citep{tran10,santos14}. In the former case, it is a factor $\sim$ 1.7
higher. For the sake of homogeneity, we have converted all the estimates to
the latter IMF, which is the same used in our simulations.}. On the other
hand, there are also examples of high redshift clusters ($z \gtrsim 2$) with
a mixed population comprising both quiescent, structurally evolved galaxies,
as well as star forming ones \citep{strazzullo13,gobat13}, or even clusters
dominated by quiescent early type galaxies \citep{tanaka13}. In any case,
{\rev there have been reports of} a reversal of the SFR-density relation,
showing increasing SFR with increasing local density at $z\gtrsim 1$, both in
the field as well as in higher density environments
\citep{elbaz07,cooper08,tran10,santos14,santos14b}.

To shed light directly on the history of assembly of galaxy clusters, it is
clear that larger samples are highly demanded at high redshift, greater than
$1 - 1.5$. Moreover, various samples should be selected by means of different
techniques, in order to capture different evolutionary stages of clusters
or proto-clusters. A widely used method to
discover high redshift clusters has been to look for the effects of their hot
gas component, namely its X-ray emission or the SZ effect it produces on the
cosmic microwave background. The former type of selection becomes rapidly
inefficient at $z \gtrsim 1.5$, due to sensitivity limitations, and both require
massive and well relaxed structures, whose number density is expected to be
rapidly declining at such early cosmic epochs. A complementary possibility,
which has been exploited in the past few years, is to preselect overdensities
of galaxies whose near infrared photometric properties are characteristic of
high redshift systems. These overdensities require later spectroscopic
confirmation. At $z>2$ most efforts have been devoted in the search of
protoclusters using  high-z giant radiogalaxies as tracers \cite[e.g.][]{hatch11,rigby14,dannerbauer14}.

A recently explored alternative to select high z clusters of galaxies has
been used by \cite{clements14}, taking advantage of the well assessed
efficiency of far-IR/sub-mm surveys in detecting high-z objects in a violent,
dust obscured, star forming phase \citep[e.g.][and references
therein]{coppin06}. These authors, following the suggestion by
\cite{negrello05},
tested the idea of exploiting  the all sky coverage of the
Planck satellite survey, in order to detect candidate clusters undergoing a pristine
and violent star forming phase. These are expected to show up as cold compact
sources, significantly contributing to the Planck number counts, which can be later
confirmed as clumps of high-z galaxies by means of higher angular resolution
maps produced by the Herschel satellite. In this first demonstrative study,
\cite{clements14} uncovered four such sources looking just at the
$\sim  90$ sq.\ deg.\ of sky observed by Herschel as part of the HerMES
survey.

The main purpose of this paper is to compare these findings with the predictions
of our high resolution simulations of the formation of massive galaxy clusters
\cite[e.g.][]{ragone13,planelles14}. To do this, we post-process the simulation
results at high $z \gtrsim 1$ with \gtd\ \citep{g3d}, a recently developed fully three dimensional
radiative transfer code, which computes the dust reprocessing of the primary photons
emitted by stellar populations (or other sources), and it has been developed
specifically to deal with the output of simulations. This post-processing is of course
required in order to properly compare with observations at IR or sub-mm wavelengths.
We discuss to what extent the spectrophotometric properties of the cluster region are
robust against reasonable variations of \gtd\ assumptions, in the various spectral regions.
We incorporate in \gtd\ a consistent treatment of the radiative effect of AGN activity
(our simulations include AGN feedback), which allows us to predict
its contribution to the emitted specific luminosity.

Besides \gtd\, a few other tools exist with similar capabilities, and have
been coupled with simulation output, e.g.: the later version of SUNRISE
\citep{jonsson10}; RADISHE \citep{chakrabart09}; ART2 \citep{li08} and SKIRT
\citep{camps14}. All but \gtd, which uses ray-tracing and finite difference, use Monte Carlo
techniques to follow the transfer of photons through the diffuse ISM, and
hence the dust re-emission. However, to the best of our knowledge, this is the first
time that this post-processing has been applied to galaxy clusters as a whole.

The paper is organized as follows: Section \ref{sec:simu} is devoted to a
brief description of the simulations set, and contains references to previous papers,
wherein all the details can be retrieved; in Section \ref{sec:g3d}
the radiative code used to post-process the simulations, \gtd, is described,
including a few modifications introduced specifically for the purposes of
this paper; the results are presented and discussed in Section
\ref{sec:results}, and summarized in the final Section \ref{sec:conclu}.

\section{The simulations}
\label{sec:simu} For this work we used simulations of 24 Lagrangian
regions extracted from a low resolution N--body simulation within a
cosmological box of $1 h^{-1} Gpc$ comoving size. We assumed a flat
$\Lambda$CDM cosmology with the following parameters: matter density
parameter $\Omega_m = 0.24$; baryon density parameter $\Omega_b =
0.04$; Hubble constant $h = 0.72$; normalization of the power spectrum
$\sigma_8 = 0.8$; primordial power spectral index $n_s = 0.96$. The
Lagrangian regions surround the most massive halos identified at $z=0$
in the parent simulation, all having virial mass\footnote{The virial
  radius and the virial mass are defined as the radius and the mass of
  the sphere encompassing a mean density equal to the over-density of
  virialization, as predicted by the spherical collapse model, and for
  the cosmology adopted in this paper \citep{bryan98}.} of at least
$10^{15}h^{-1}M_\odot$ \citep[e.g.][]{bonafede11}. Initial conditions
for the hydrodynamical simulations have been created by increasing
mass resolution within such regions, and adding the corresponding
high-frequency Fourier modes from the linear power spectrum of the
adopted cosmological model. In The mass of DM particles is $8.47
\times 10^8 h^{-1} \mbox{M}_\odot$, and the initial mass of each gas
particle is $1.53\times 10^8 h^{-1} \mbox{M}_\odot$.

Our simulations were performed using the TreePM-SPH {\small GADGET-3}
code, an improved version of {\small GADGET-2} \citep{springel05}. The
force accuracy, in the high resolution regions, is set by $\epsilon =
5 h^{-1}$ kpc for the Plummer-equivalent softening parameter, fixed in
comoving units at redshifts $> 2$ and fixed in physical units in the
redshift range $2 \le z \le 0$. When computing hydrodynamical forces,
the minimum value attainable for the Smoothed Particle Hydrodynamics
(SPH) smoothing length of the B-spline kernel is set to half of the
gravitational softening length.

The simulations used in this work include gas cooling, star formation,
supernova feedback and AGN feedback. {\rev We adopted a Chabrier Initial Mass Function
\citep[IMF,][]{chabrier03}}. For a detailed description of the
sub-resolution models we refer the reader to \cite{ragone13} and
\cite{planelles14}. In particular, the former paper contains a discussion of
the prescriptions for the BH feedback, which is based on the simple recipe
put forward by \cite{Springel05b} and adopted in many simulations, woth a
few important modifications.  These were necessary because the latter model
was thought and calibrated for non-cosmological high resolution simulations
of merging galaxies. {\rev It is now well recognized that the sub-resolution
prescriptions are sensitive to resolution, and as such they generally require
re-calibration or even a deeper rethinking when the resolution is changed \cite[see
e.g.][]{crain15}. Indeed, we found that the recipes proposed by
\cite{Springel05b} for the BH feedback leads, in the context of our
significantly lower resolution cosmological simulations, to several unwanted
and misleading effects, discussed in \cite{ragone13}, such as unrealistic
merging of distant SMBH particles or losses of the energy produced by
accretion. Our approach there has been to introduce the minimal changes
required to avoid unreasonable results.}

Since some concepts of this modelling are used in Section
\ref{sec:g3d}, {\rev to help the reader we provide here a brief summary.}
The BHs are represented by means of collision-less particles, subject only to
gravitational forces, and growing by accretion and merging. The
accretion rate is given by the minimum between a Bondi accretion rate,
modified by the inclusion of a multiplicative factor, and the
Eddington limit. The former is loosely thought of as providing an
estimate of how the gas available for accretion scales with the
conditions in the BHs surroundings, while the Eddington limit ensures
that the produced radiation pressure does not overcome gravity. When
two BH particles are within the gravitational softening and their
relative velocity is smaller than a fraction 0.5 of the sound velocity
of the surrounding gas, we merge them. The accretion onto SMBHs
produces an energy determined by a parameter $\epsilon_r=0.2$, giving
the fraction of accreted mass converted to energy. Another parameter
$\epsilon_f=0.2$ defines the fraction of this energy that is thermally
coupled to the surrounding gas. As usual, we calibrated these
parameters in order to reproduce the observed scaling relations of
SMBH mass in spheroids at z=0.

For the analysis presented in this paper, we identified cluster
progenitors at several redshifts $z\le 3$. In particular, our sample has a median
virial mass of $8 \times 10^{13} \hmass$ and $2 \times 10^{14} \hmass$
at $z=2$ and 1, respectively.  The corresponding median SFRs within
the virial radii are of 800 and 500 \msunyr.

\subsection{Cluster selection and initial conditions}
\label{sec:clusel}
As described in full details in \cite{bonafede11}, the 24 most massive
clusters, in terms of {\rev the mass assigned by the adopted cluster identification
algorithm {\sl Friend of Friends} (FOF; Davis et al. 1985)}, have been selected in the parent
simulations at z=0. The re-simulations at higher resolution have been carried
out using the {\it Zoomed Initial Conditions} technique \citep{tormen97}.
The HR regions allow us to identify other 50 less massive and
interesting clusters, uncontaminated by low resolution particles, which have
been studied in several papers \citep[e.g.][]{ragone13,planelles14}. However,
here we focus only on the 24 originally selected ones, because they
constitute a statistically well defined sample, whose final virial mass is in
the range between $\simeq$ 1 and $3 \times 10^{15} \hmass$ (see table 1 in
Bonafede et al.\ 2011). Our sample has a median virial mass of  $8 \times 10^{13} \hmass$
and $2 \times 10^{14} \hmass$ at redshift 2 and 1 respectively, and the
corresponding median SFRs within the virial radius are of 800 and 500 \msunyr.

\section{\gtd\ and its modifications}
\label{sec:g3d}

In order to perform radiative transfer calculations in the region of
simulated clusters, we use \gtd\ \citep{g3d}, a recently developed fully three
dimensional radiative transfer code, which computes the dust reprocessing of
the primary photons emitted by stellar populations (or other sources). \gtd,
while largely based on the formalism of the widely used model GRASIL
\citep{silva98,granato00}, has been specifically designed to be
applied to systems with arbitrary geometry, in which radiative transfer
through dust plays an important role, such as galaxies or interesting regions
identified in the output of hydrodynamical galaxy formation codes. With
respect to the already published version, we have introduced here a few
modifications to adapt it to the output of our version of GADGET-3, in
particular for what concerns the radiative effect of AGN activity.
The main features of \gtd\ are summarized below, while we refer the
reader to \cite{g3d} for all the details. A description of
the modifications introduced for the purposes of this paper follows this summary.

A somewhat overlooked point is that any sensible computation of radiative
transfer from the output of cosmological simulations, inevitably
requires the introduction of a few more free or uncertain parameters, with
respect to those already demanded by the treatment of baryon physics in the
simulation itself. These are related to the sub-resolution astrophysics, in
particular that concerning the molecular clouds (MCs). Since MCs are the sites
of star formation and massive stars spend part of their lives within or close
to them, it is expected and well established that in star forming
systems a significant fraction of dust reprocessing occurs precisely in MCs.
This fraction is an increasing function of the specific star formation
activity, and of the total reprocessing itself \citep[e.g.][]{silva98,granato00}.
Current cosmological hydro-dynamical simulations that follow galaxy
formation are only beginning to resolve MCs (we remind the reader that the typical sizes
and masses of giant MCs are of the order of $\sim 10-20$ pc and $10^5-10^6$
M$_\odot$ respectively),  for zoom-in simulations of single galaxies
\citep[e.g.][]{hopkins14}, but their high computational cost makes them un-doable
as yet for simulations on cluster scales. Therefore some further
sub-resolution modelling is required to cope with dust reprocessing.

%Unfortunately, current cosmological hydro-dynamical codes that follow galaxy
%formation are very far from resolving MCs (we remind that the typical sizes
%and masses of giant MCs are of the order of $\sim 10-20$ pc and $10^5-10^6$
%M$_\odot$) therefore some further sub-resolution modelling is required.

In particular, in \gtd\ the ISM is divided into two components, the molecular
clouds (MCs) and the diffuse {\sl cirrus}. In order to calculate the mass in
the form of molecular clouds it is assumed that unresolved gas densities at
any point of the simulated volume follow a log-normal probability
distribution function (PDF), The mean of the PDF is given by the local gas density and
dispersion $\sigma$ (a free parameter), as suggested by small scale ($\sim$ 1
kpc) simulations. Then, the local contribution to the molecular fraction is
given by the fraction of the PDF above a threshold density, $\rho_{MC,
thres}$. The two parameters introduced so far, which control the molecular
gas fraction are, $\rho_{MC, thres}$ and $\sigma$, may reasonably range
from about 0.3 to 3 M$_\odot$ pc$^{-3}$ and from 2 to 3 respectively.

Once the molecular fraction at any location of the system is calculated, \gtd\
takes into account the age-dependent dust reprocessing of stellar
populations. This age dependency arises from the fact that younger stars are associated with
denser ISM environments (note that GRASIL was the first model to take it into account).
This is obtained assuming that stars younger than a certain time $t_0$ (a
further parameter) are enshrouded within molecular clouds, characterized by
their mass $M_{MC}$ and radius $R_{MC}$. Then, the radiative transfer is
treated separately in the two components with the required accuracy. We
explicitly note that, even though we apparently introduced here two more
parameters $M_{MC}$ and $R_{MC}$, what actually matters for the radiative
transfer computation is the ratio $M_{MC}/R_{MC}^2$, which therefore should
be regarded as the only new parameter. A detailed non-equilibrium calculation
for dust grains smaller than a given radius ($150$ \AA\ in this work) and for
polycyclic aromatic hydrocarbons molecules (PAHs) is performed. Their emission
is usually very important at $\lambda \lesssim 30 \ \mu$m in the cirrus component.

\gtd\ has a general applicability to the outputs of either Lagrangian or
Eulerian hydrodynamic codes. The first applications of the code have been done
interfacing it with the output of the P-DEVA and {\tt GASOLINE} SPH codes
\citep{g3d,obreja14}.

\begin{figure}
\hspace{-1cm}
\includegraphics[width=8.5cm, height=7.5cm]{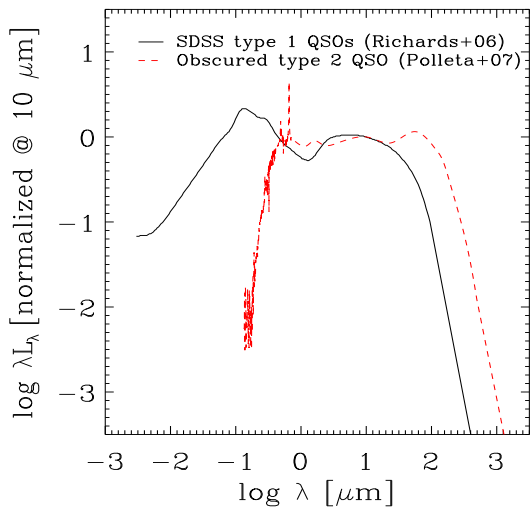}
\caption{The SED templates adopted in this work to describe the emission of SMBH
particles.
In our standard computations we used the mean SED of SDSS type 1 quasars by
Richards et al.\ 2006 (black solid line). To check the stability of our results
on the IR properties
of clusters, we used the very different typical SED of an obscured
type 2 QSO, reported by Polletta et al.\ 2007 (red-dashed line).
}
\label{fig:templates}
\end{figure}

The main modification of \gtd\ introduced in this work has been to account
for the radiative effect of SMBH particles present in the simulation, which
provide the AGN feedback as mentioned in Section \ref{sec:simu}. To do this,
we assume that the radiation emitted by the SMBH particles is distributed
according to a template SED, properly normalized by the bolometric
luminosity. The latter is computed from the accretion rate and from the
radiative efficiency, as $L_{bol}=\epsilon_r \dot M c^2$. We remind the
reader that the former quantity is predicted by the simulation for each SMBH
particle, while the efficiency $\epsilon_r$ is a parameter of the simulation.
It is set to $0.2$ in the runs used in this work \citep[see][for
details]{ragone13}. As for the SED, we adopted in our standard computation
the mean SED of SDSS quasars computed by \cite{richards06}, plotted in Figure
\ref{fig:templates}. It is observationally known and theoretically expected
that AGN are characterized by a substantial anisotropy of their emission,
giving rise to the broad dichotomy between type 1 and type 2 AGN. This is
likely related to a preferential axis in their central engine \citep[for a
recent review see][]{hoenig13}. However, we neglect this complication here.
This is owing to two reasons. The former is that our simulations do not give
any prediction for the AGN preferential axis, which is possibly related to
the SMBH spin. The second and most reassuring one is that we checked that
different choices for the assumed template SED yields only very small
differences in the predicted IR properties of clusters, which become usually
negligible at $\lambda \gtrsim 100 \mu$m (see Section \ref{subsec:eff_SMBHs}
and Figure \ref{fig:cfrt_sed_radoff}). This holds true even for the totally
opposite case of assigning to all SMBH particles the typical SED of obscured
type 2 QSO, such as that reported by \cite{polletta07}, also shown in Figure
\ref{fig:templates}.

We note that these observational templates include by construction the
reprocessing by dust in the region close to the SMBH, namely that which has
been long since ascribed to a torus-like structure. The latter has been
invoked to explain the observational differences between type 1 ({\sl
un-obscured}) and type 2 ({\sl obscured}) AGNs
\citep[e.g.][]{pier93,granato94,efsta95}. This is actually what we need.
Indeed, even though little can still be safely concluded on the detailed
geometrical properties of these structures \cite[e.g.][and references
therein]{hoenig13}, even the more extended models do not consider torii
larger than $\sim 200 \left(L_{AGN}/(10^{46} \mbox{erg s}^{-1}
)\right)^{1/2}$ pc \citep{granato97}, which is far below the resolution of
any cosmological simulation.

We assumed that SPH gas particles contain dust only when their temperature is
lower than a certain threshold, that in the present work we set to $10^5$
Kelvin. However, the results are very weakly dependent on this value. For
instance, we verified that the predicted SEDs of clusters are almost
indistinguishable if the threshold is increased or decreased by a factor
$\sim 10$. The maximum difference occurs at around the peak of dust emission
$\sim 100 \, \mu$m and in the far UV, and is $\lesssim 10$ \%. The dust to gas
ratio $\delta$ of gas particles colder than the threshold temperature was assumed
to be proportional to their metallicity, with a proportionality constant
calibrated to get the standard galactic value $1/110$ at solar metallicity,
i.e.\ we set $\delta=Z/(110 \, Z_\odot)$. This corresponds to assuming that
about 50\% of metals are locked in dust grains.

\begin{figure*}
\hspace{-1cm}
\includegraphics[width=9.5cm]{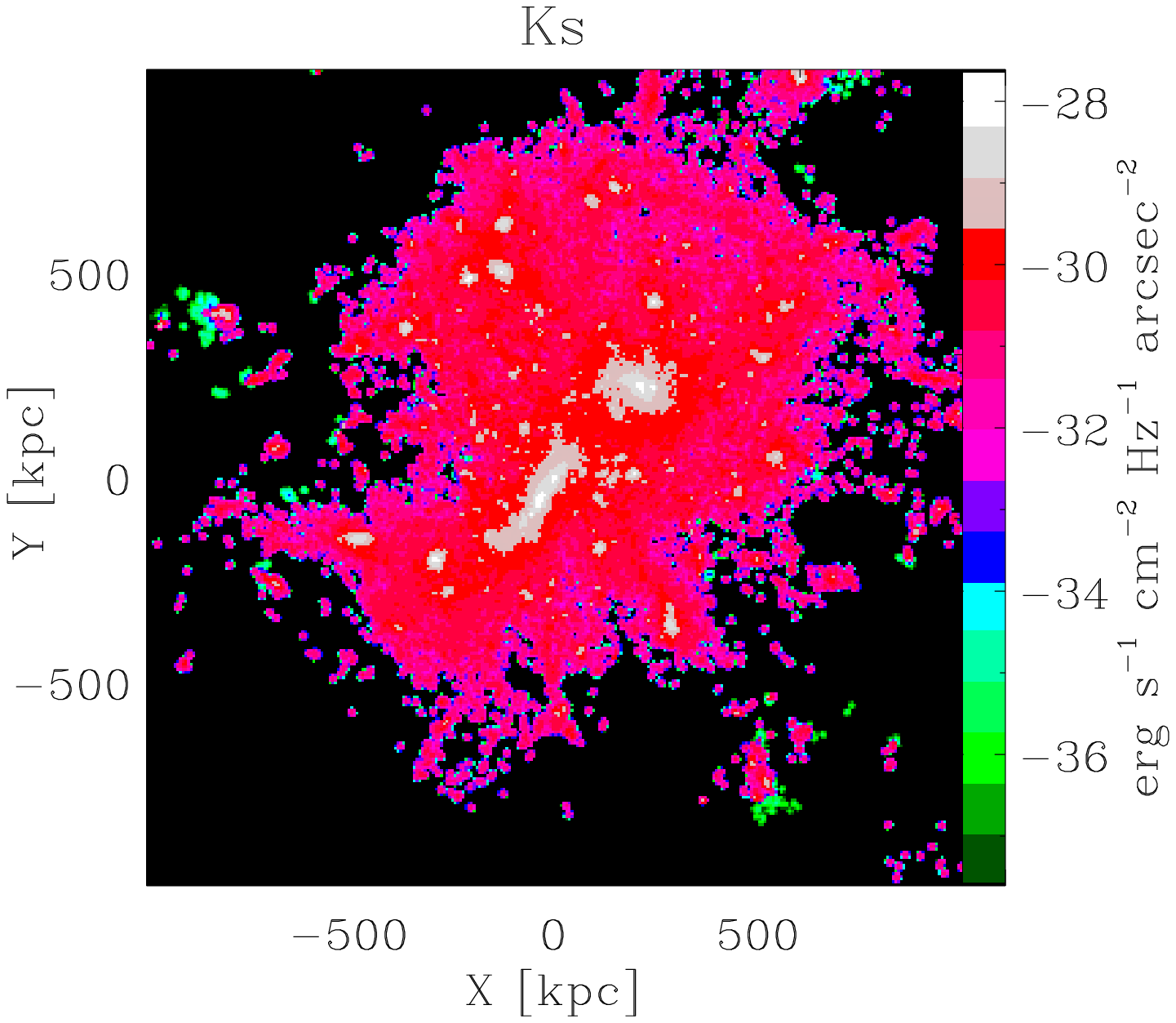}
\hspace{-1.2cm}
\includegraphics[width=9.5cm]{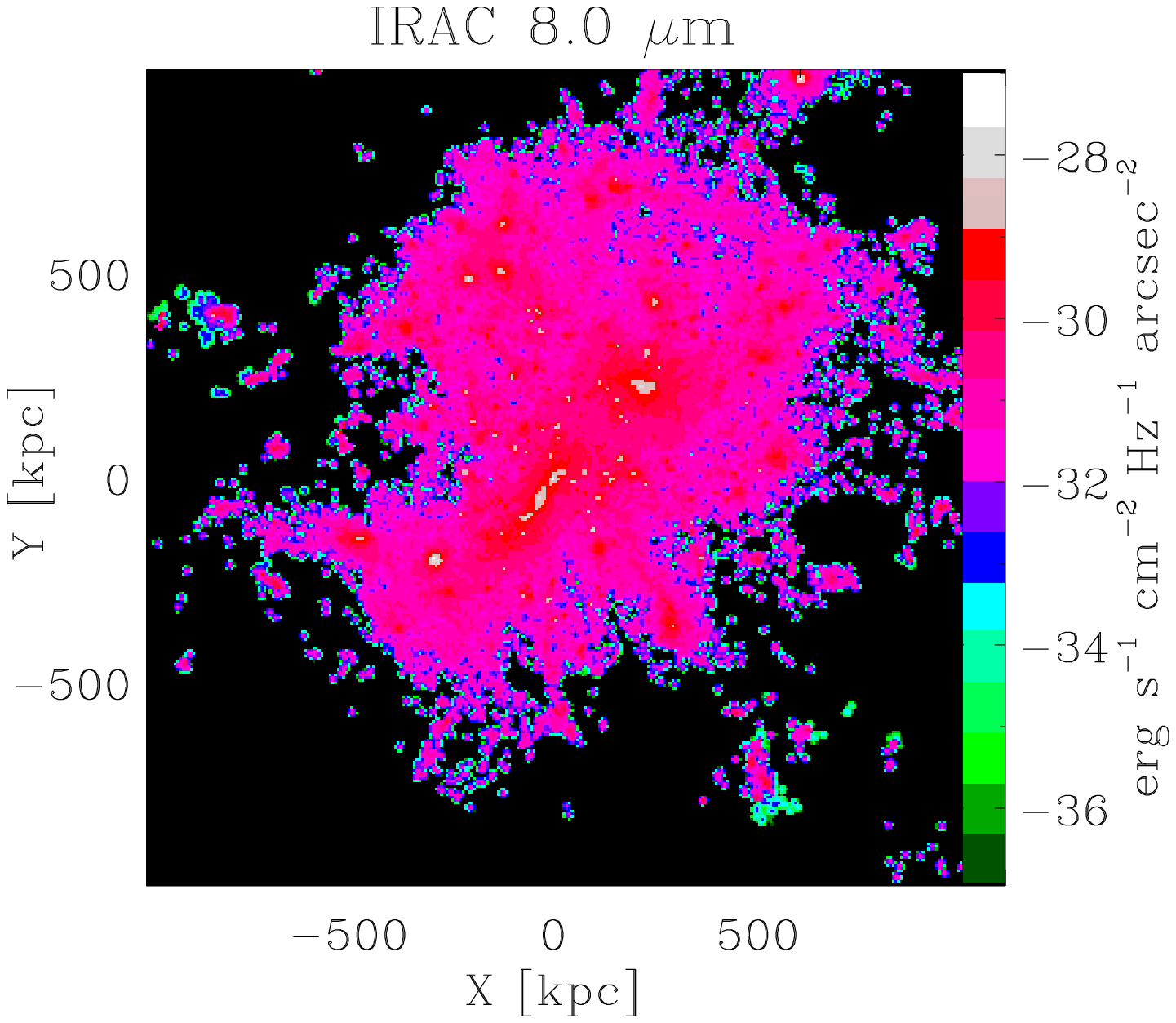}

\vspace{0.3cm}
\hspace{-1cm}
\includegraphics[width=9.5cm]{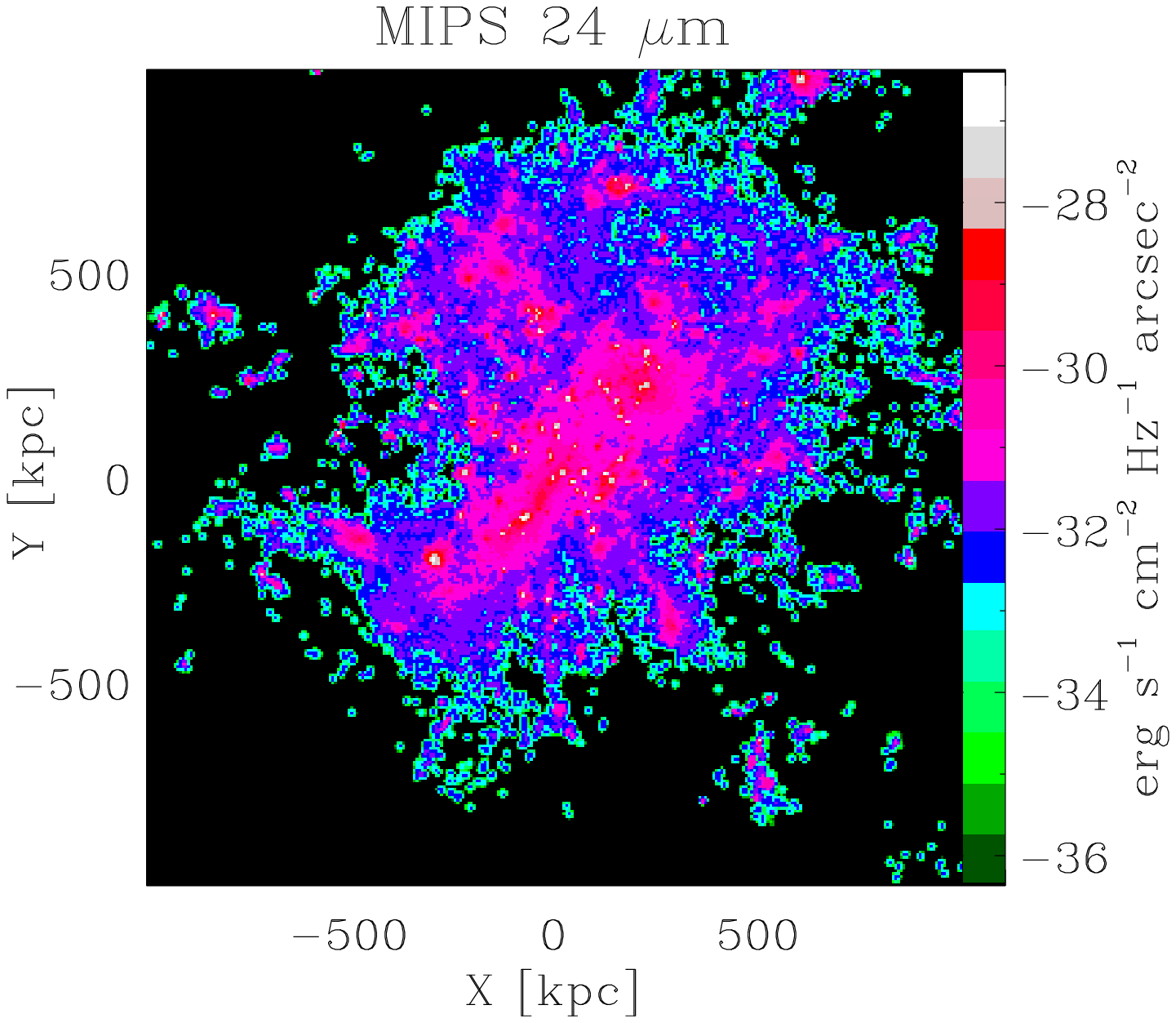}
\hspace{-1.2cm}
\includegraphics[width=9.5cm]{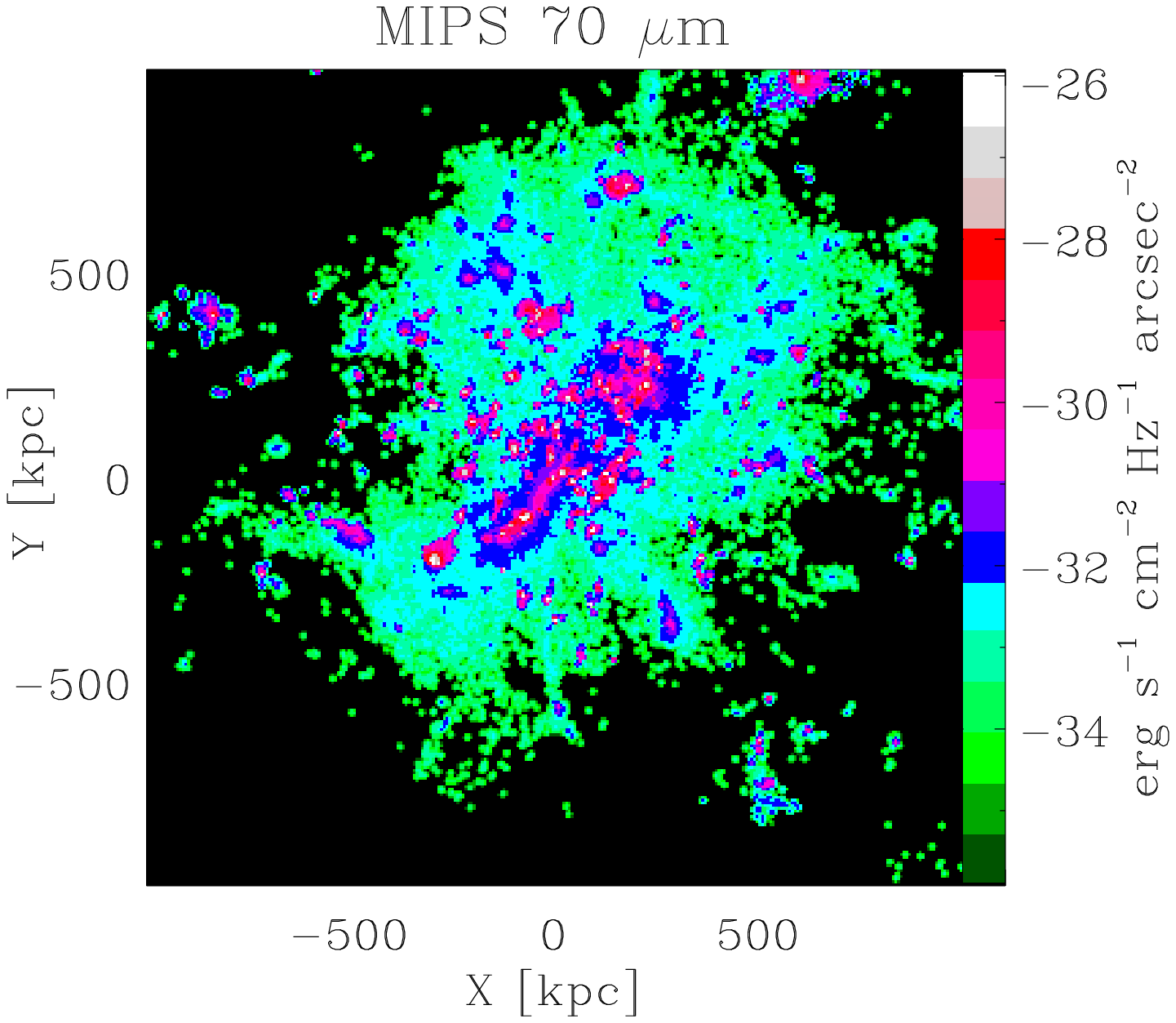}

\vspace{0.3cm}
\hspace{-1cm}
\includegraphics[width=9.5cm]{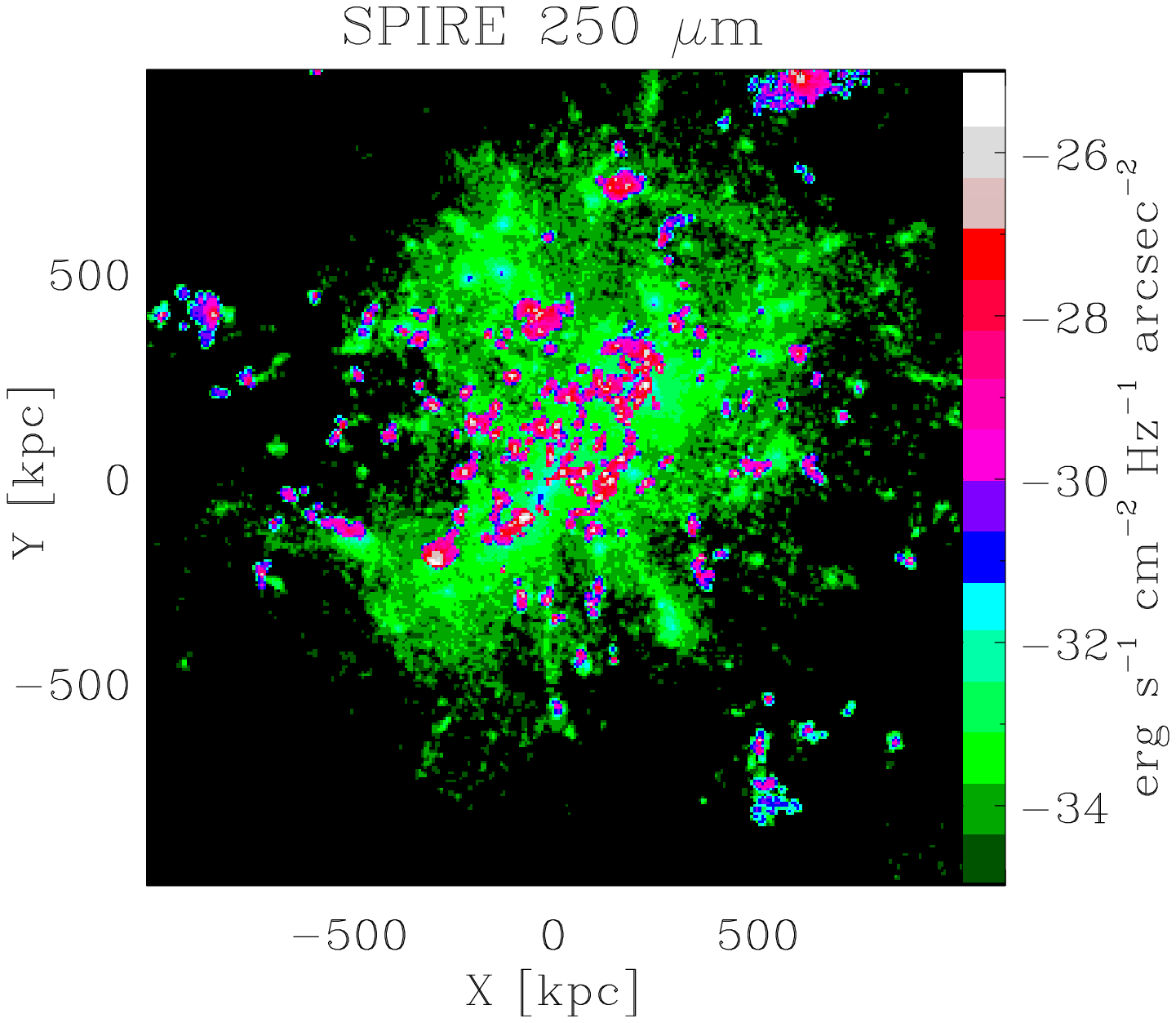}
\hspace{-1.2cm}
\includegraphics[width=9.5cm]{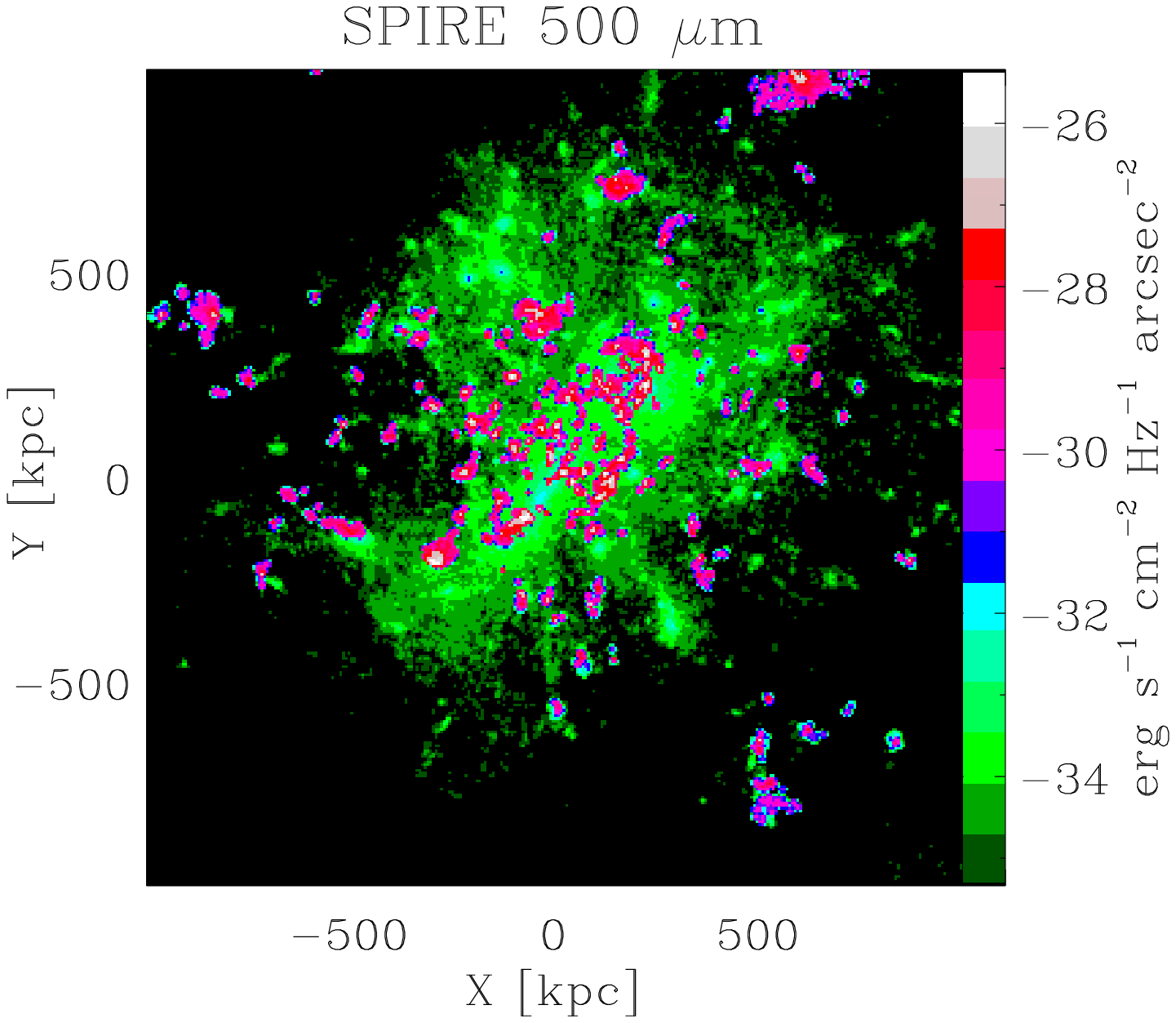}
\vspace{0.5cm}
\caption{Examples of images of a cluster region at z=1 produced by \gtd\, in various NIR to sub-mm bands.
The physical size of each panel is 2000 kpc, close to the Planck HFI beam at that redshit.
No telescope effects (like point spread functions, pixel sizes, etc.) have been taken into account.}
\label{fig:maps_z1}
\end{figure*}

\begin{figure*}
\hspace{-1cm}
\includegraphics[width=9.5cm]{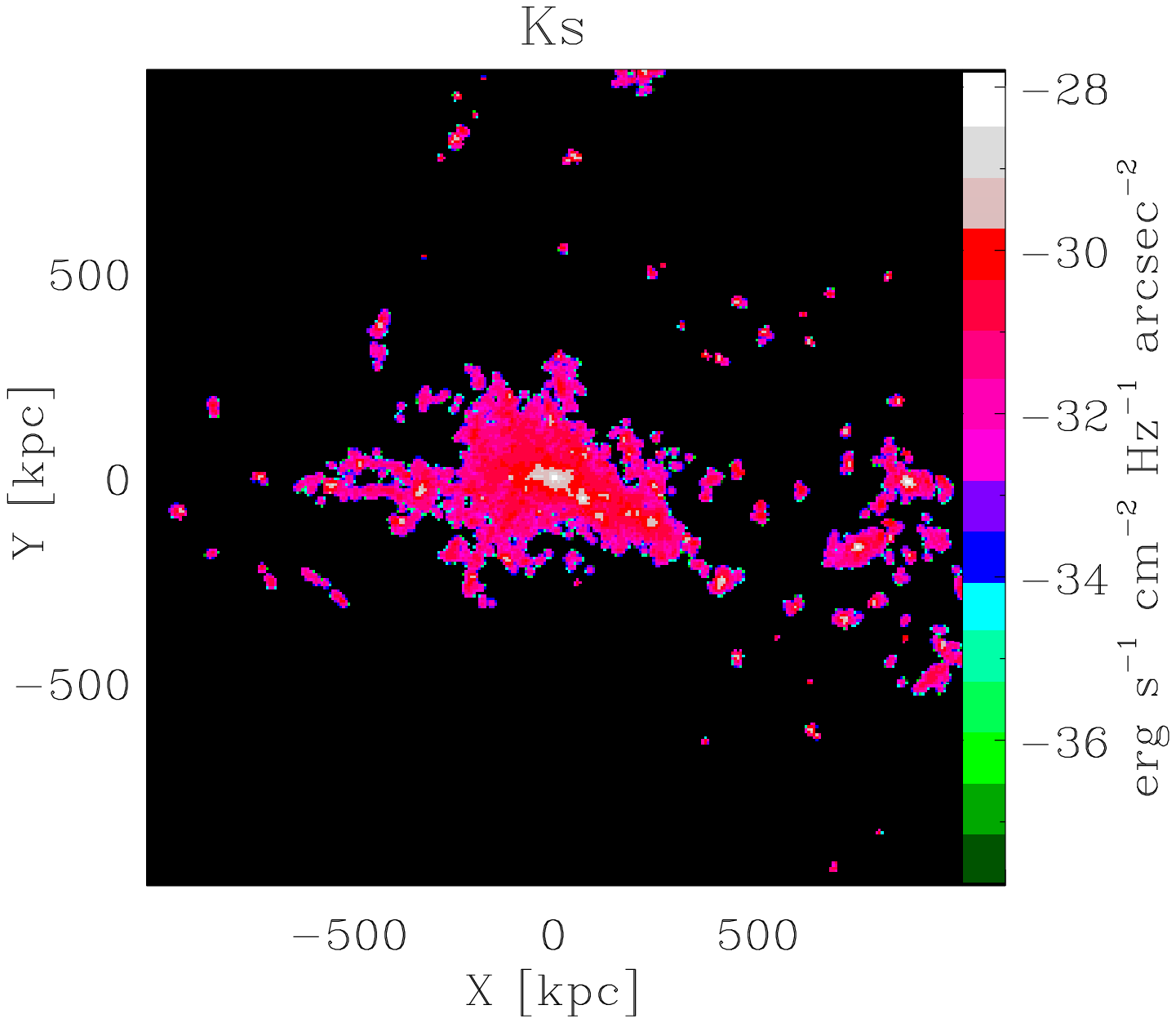}
\hspace{-1.2cm}
\includegraphics[width=9.5cm]{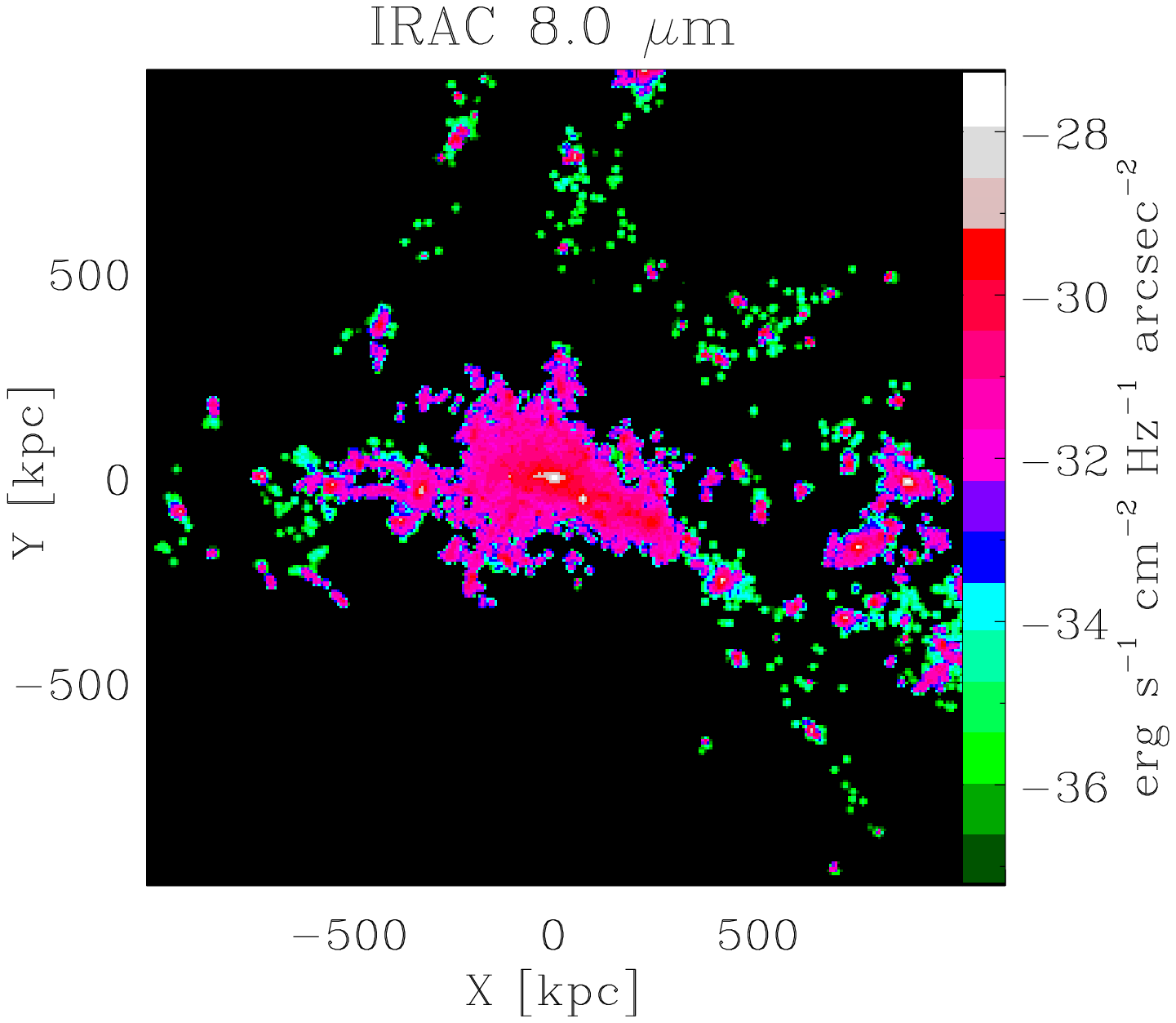}

\vspace{0.3cm}
\hspace{-1cm}
\includegraphics[width=9.5cm]{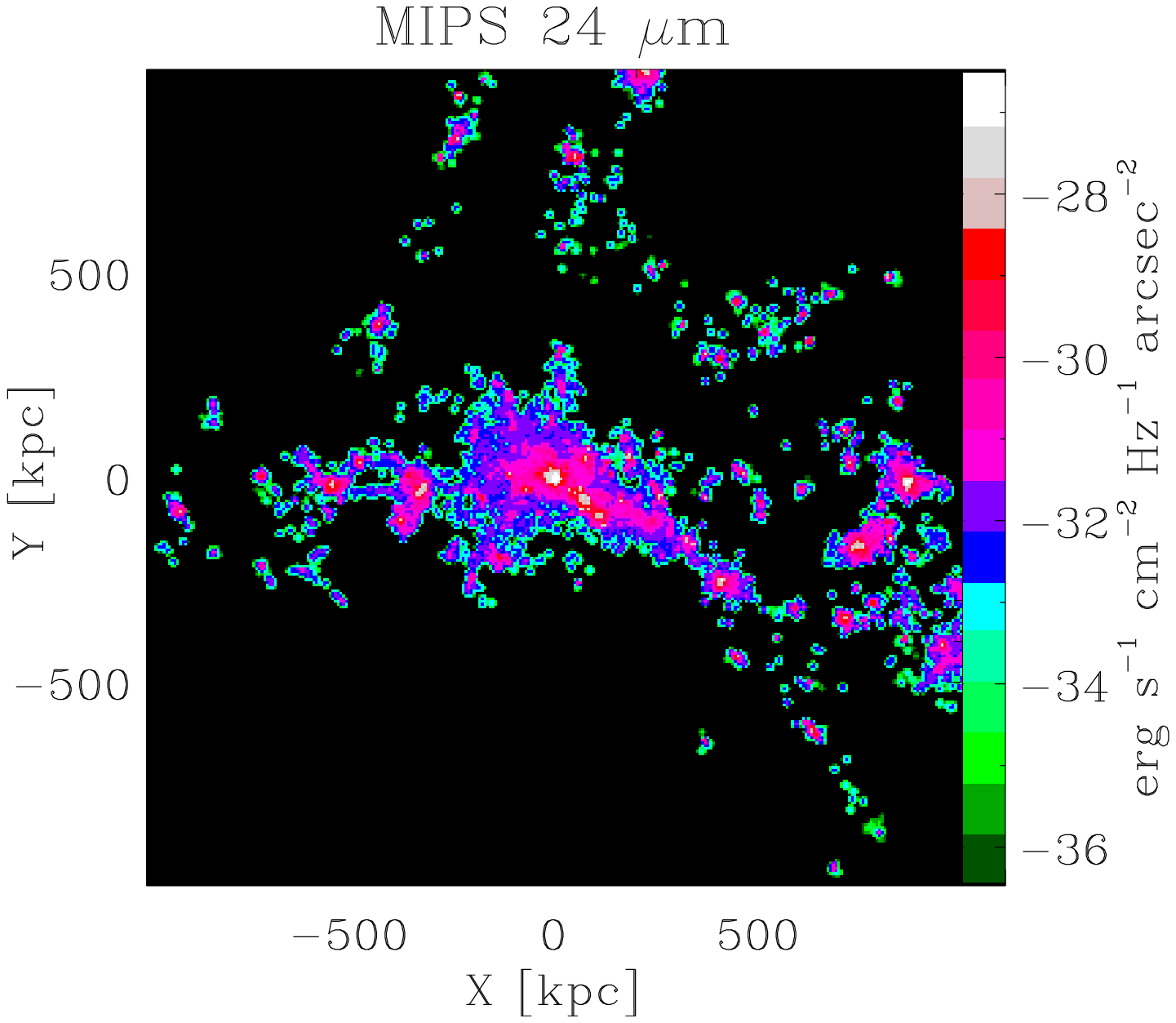}
\hspace{-1.2cm}
\includegraphics[width=9.5cm]{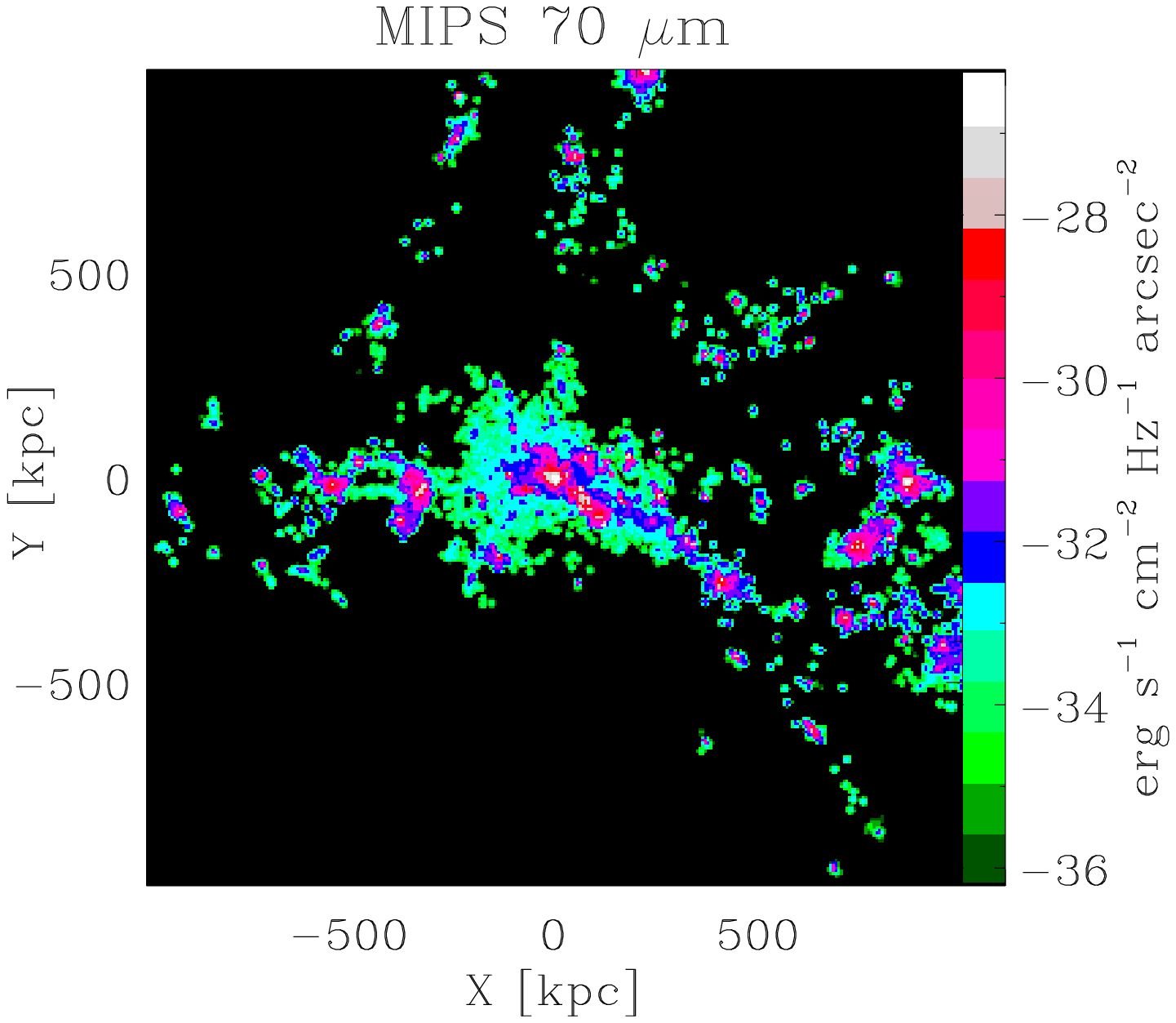}

\vspace{0.3cm}
\hspace{-1cm}
\includegraphics[width=9.5cm]{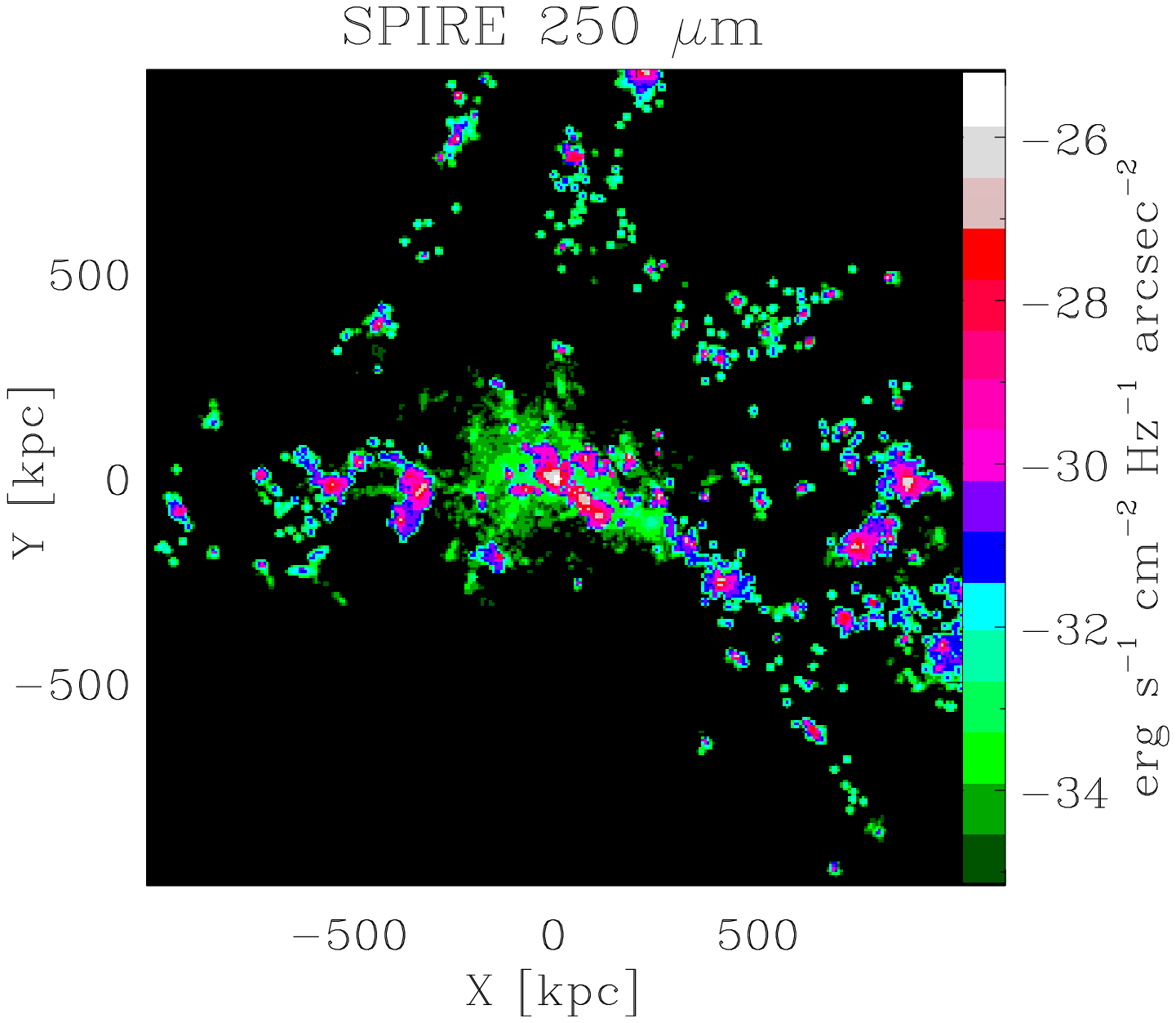}
\hspace{-1.2cm}
\includegraphics[width=9.5cm]{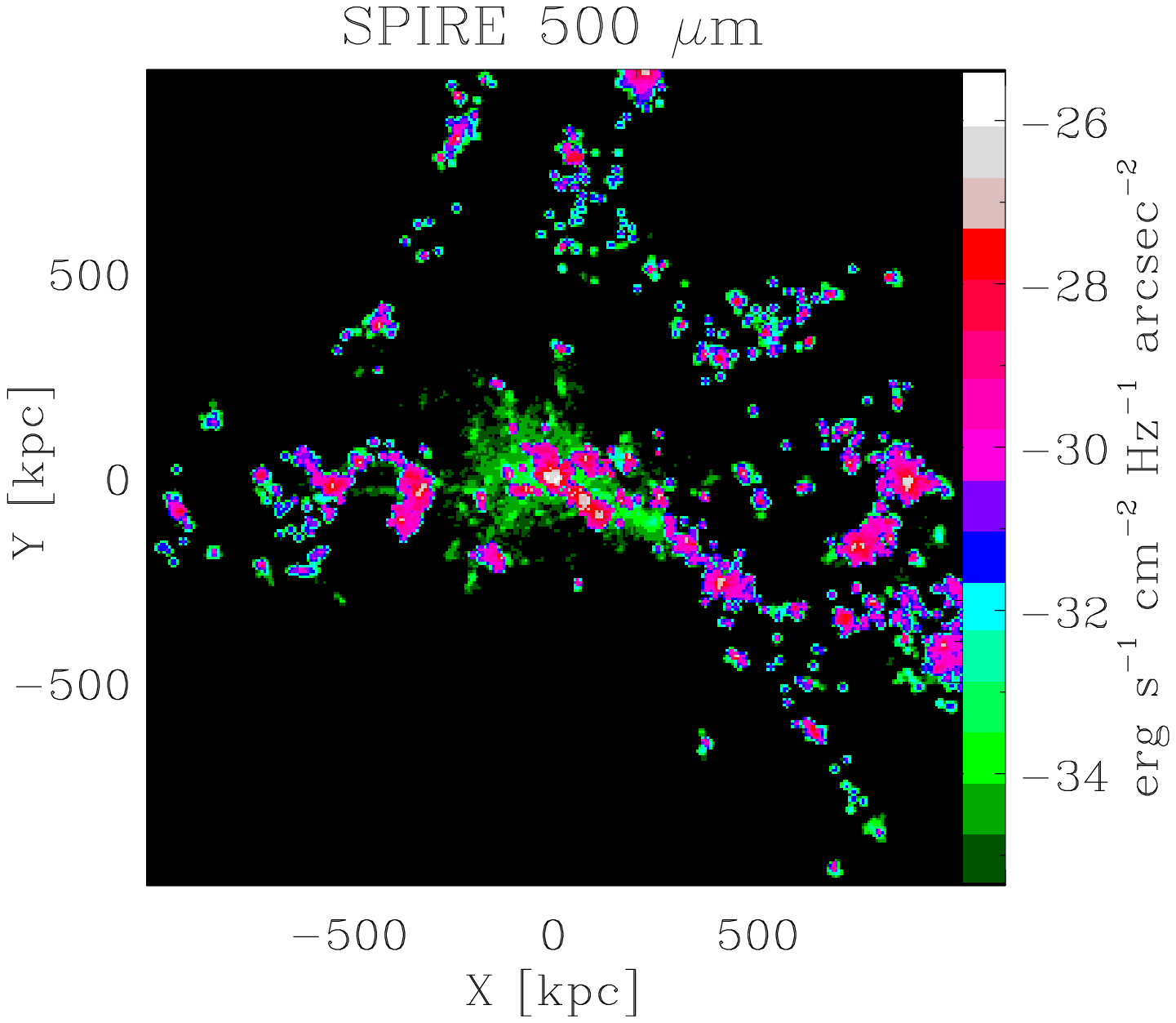}
\vspace{0.5cm}
\caption{Same as Fig.\ \ref{fig:maps_z1} but at z=2}
\label{fig:maps_z2}
\end{figure*}

\section{Results}
\label{sec:results}

\begin{table*}
  \centering
  \begin{tabular}{| l | l | l | l | }
     \hline
     % after \\: \hline or \cline{col1-col2} \cline{col3-col4} ...
     Parameter & Adopted value & Reasonable range & Short description\\
     \hline
     $t_0$ & 6 Myr & 1 to 30 Myr & Escape timescale of stars from parent MCs\\
     $M_{MC}/R^2_{MC}$  & $5 \, 10^5 \, \mbox{M}_\odot/ (10 \, \mbox{pc})^2$  & $10^5 \, \mbox{ to } \, 10^6 \, \mbox{M}_\odot/ (10 \, \mbox{pc})^2$  & Determines MCs optical depth\\
     $\rho_{MC, thres}$ & $1\,  \mbox{M}_\odot/\mbox{pc}^3$  &  0.3 to 3 M$_\odot/\mbox{pc}^3$ & Threshold density for gas to be
     considered in MC phase\\
     $\sigma$ & 2.5 & 2 to 3 & Dispersion of the sub-resolution PDF of gas densities \\
     \hline
   \end{tabular}
  \caption{Parameters introduced by \gtd\ computation.}
  \label{table:pars}
\end{table*}

\begin{figure*}
\hspace{-1cm}
\includegraphics[width=8.5cm, height=7.5cm]{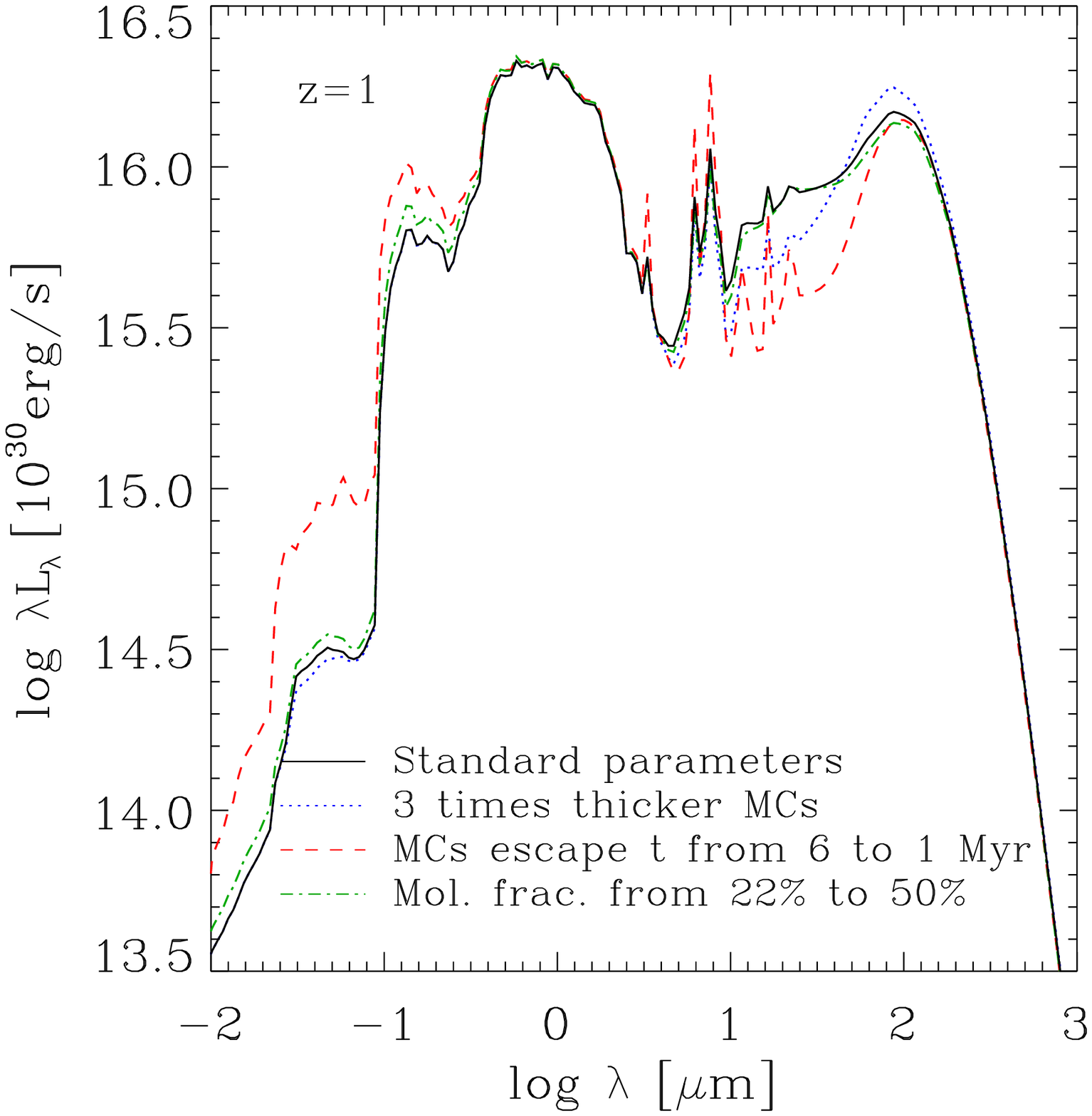}
\includegraphics[width=8.5cm, height=7.5cm]{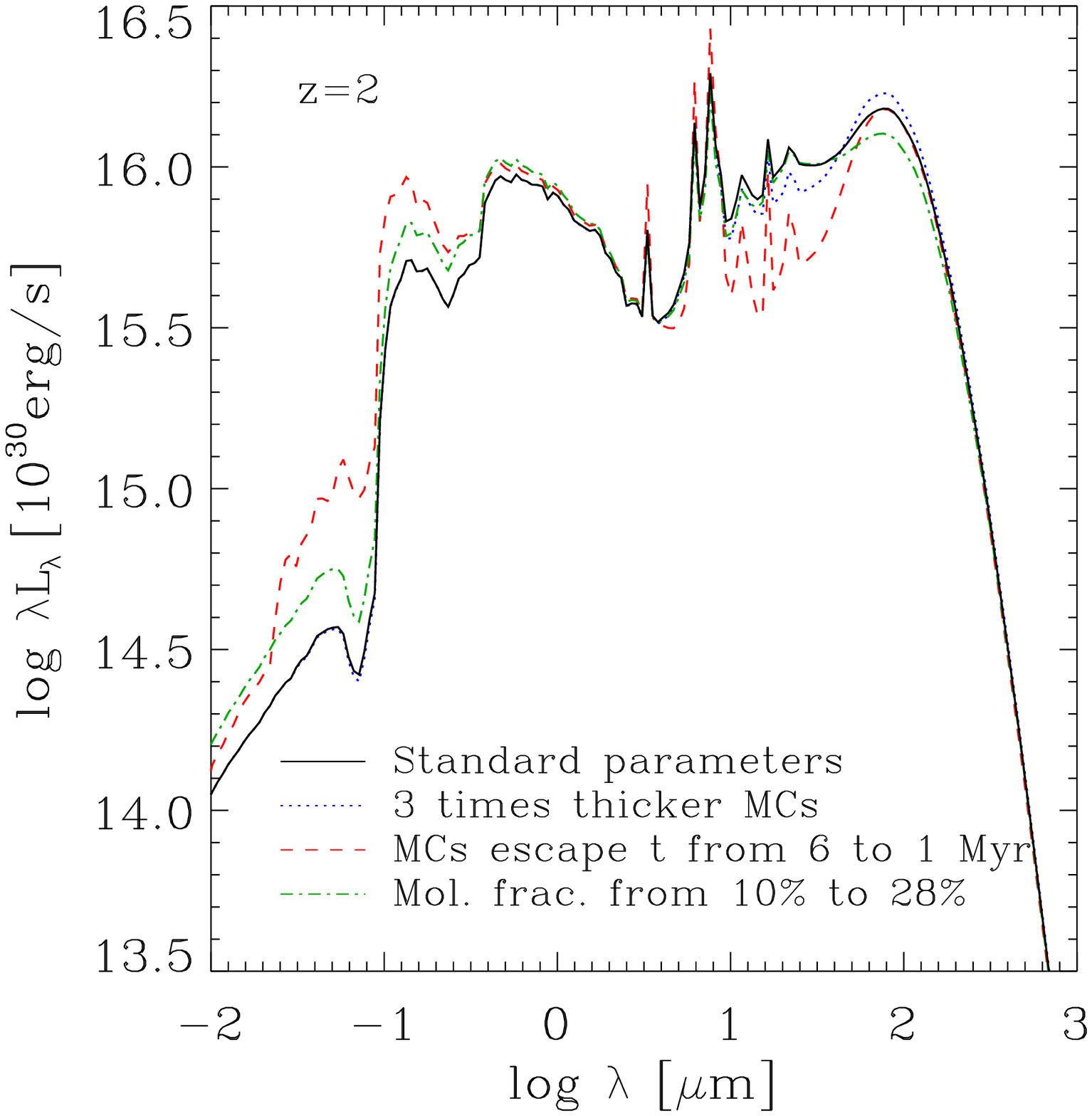}
\caption{Comparison of the rest frame SED obtained for one cluster at z=1 and 2,
with different choices of \gtd\ parameters.
Solid-black: standard values; blue-dotted: increasing the mass of individual
MCs (and thus
their optical depth) by a factor 3; red-dashed: decreasing the escape
timescale of
stars from MC by a factor 6 (from 6 to 1 Myr); green-dot-dashed: decreasing the
threshold for gas to be in the dense MC phase by a factor 10, so that the
molecular gas fraction increases from about 22\% to about 50\% at z=1 and from
about 10\% to 28\% at z=2.
In any case, the SEDs are very little affected
above $\lambda \gtrsim 100\ \mu$m, which is the spectral region on which we compare
with observations in this paper. Note also that the parameter variations have been adopted to
exacerbate the effects, but often leads to somewhat unrealistic values.
}
\label{fig:cfrt_sed}
\end{figure*}

\begin{figure*}
\hspace{-1cm}
\includegraphics[width=8.5cm, height=7.5cm]{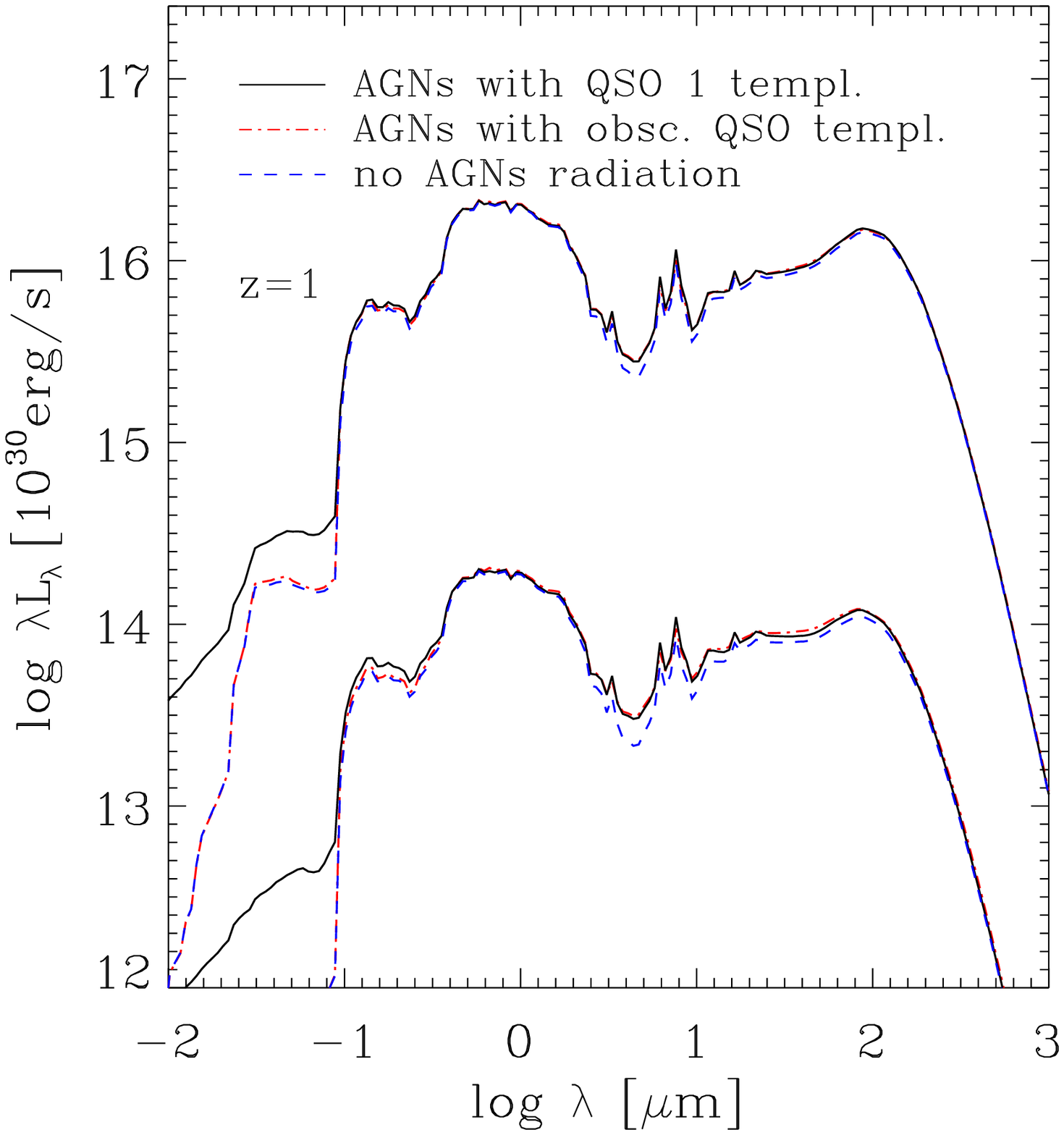}
\includegraphics[width=8.5cm, height=7.5cm]{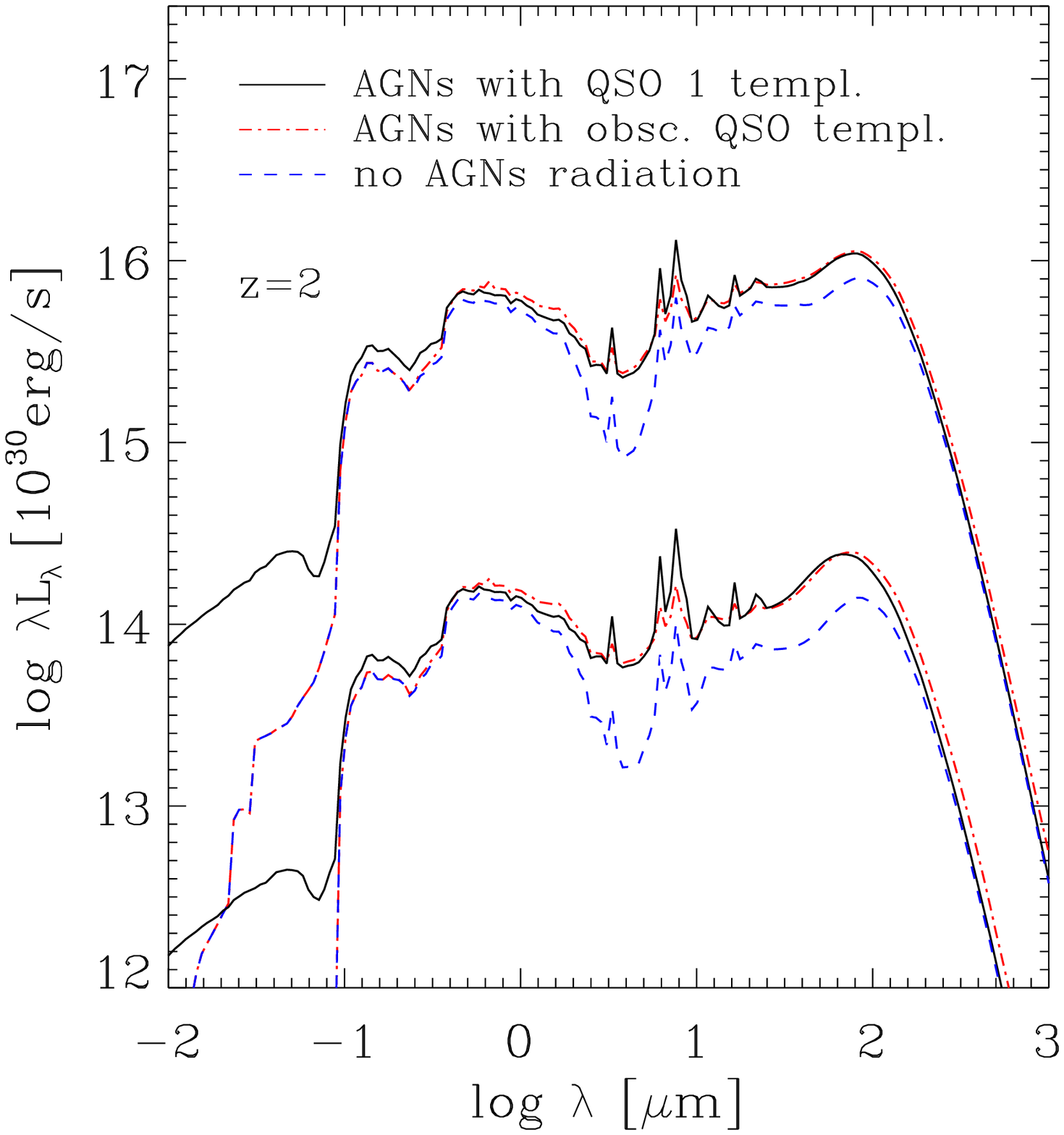}
\caption{Comparison of the rest frame SED obtained for two clusters at z=1 and
2 by including or neglecting the radiative effect of AGNs, or by adopting
different templates for it. The three lower lines have been artificially displaced
by a factor ten, and refer to a rare case ($\sim 10 \% $ at z=2,
but never at z=1) in which the differences are most prominent (at z=2), while
the three upper lines refer to the same cluster considered in
Fig.\ref{fig:cfrt_sed},
which is more typical.
Solid-black: including the radiative effect of AGNs with our standard
template, namely the mean SED of SDSS quasars by Richards et al.\ (2006);
blue-dashed: neglecting the radiation emitted by AGNs; red-dot-dashed:
including the radiative effect of AGNs, but using the template by Polletta et
al.\ (2007) for heavely obscured type 2 QSOs.}
\label{fig:cfrt_sed_radoff}
\end{figure*}

The outputs of \gtd\ are mock images of the portion of the simulated box in
which the radiative transfer has been performed, and SEDs of the radiation
coming out from the same box. We have processed  boxes of physical
size 2 Mpc, encompassing each of the 24 clusters, at several redshift between
0.5 and 3. Figures \ref{fig:maps_z1} and \ref{fig:maps_z2} show examples of
such images, in several interesting band-passes, for one of the clusters at
redshift 1 and 2 respectively. The physical size of each panel is 2000 kpc,
which at these redshift is close to the Planck HFI beams $\sim 5$ arcmin, and
is larger than the virial diameter of the clusters at both redshifts,
typically by a factor $\sim 2$ and 5 respectively. No telescope effects
(like point spread functions, pixel sizes, etc.) have been taken into
account. In this Section, we will show how the predicted SEDs arising from
the same region depend on \gtd\ parameters and other assumptions, and then we
will discuss some implications for star forming cluster searches at sub-mm
wavelengths.

%{\rev (or 2880 and 4320 $\Kpc$ comoving at z=1 and 2)}

\subsection{Dependence on \gtd\ parameters and assumptions}
\label{subsec:dep}

To summarize, the parameters introduced by the radiative transfer
calculations performed with \gtd\ are (i) the timescale for newly born stars
to get rid of the parent MC, $t_0$; (ii) the ratio between the mass of MCs
and the square of their radius $M_{MC}/R^2_{MC}$, which in conjunction with
the dust to gas ratio $\delta$ determines the optical depth of the MC, $\tau
\propto \delta \, M_{MC}/R^2_{MC}$; we remind you that $\delta$ is set by the
local gas metallicity, see end of Section \ref{sec:g3d}; (iii) the threshold
density for gas to be in the MC dense phase $\rho_{mc, thres}$ and (iv) the
dispersion $\sigma$ of the sub-resolution PDF of gas densities. The latter
two determines the fraction of gas in molecular form. Our adopted standard
values for these quantities are reported in Table \ref{table:pars}, together
with their reasonable ranges \citep[see][and references therein]{g3d}. Fig.
\ref{fig:cfrt_sed} shows the SED of the same region of Fig.
\ref{fig:maps_z1}, computed under large variations of these parameters, with
respect to our adopted standard values. These variations are thought to
exacerbate the effects.
Despite this, in the spectral region above $\sim 100\ \mu$m rest frame, which
is that required to compare with the observations discussed in this paper,
the effects are small, and become negligible above $\sim 200\ \mu$m. By converse
at shorter IR wavelengths, where the contribution of MC becomes important,
the consequences of these different choices of parameters may be
significative, and would ask for careful evaluation. In particular,
for $10 \lesssim \lambda \lesssim 40 \ \mu$m they may amount to a factor $\sim 2$.

Besides the uncertainties in \gtd\ outputs arising from the choice of its
explicit parameters, there are also those related to the adopted optical
properties of dust grains, which are by far less understood and
predictable than commonly assumed \citep[for a recent review see][]{jones14}.
In this work we retain the same dust mixtures for the MCs and for the
cirrus as in \cite{g3d}, which have been calibrated to reproduce
the average properties of local galaxies. As for computing dust emission, the
most delicate region is that below $\sim 30\ \mu$m, where the contribution
from small thermally fluctuating grains and PAHs becomes important,
and relatively minor variations just in the adopted size distribution of grains may
produce large variations. This kind of uncertainty is expected to increase
with redshift, and considering environments very different from those in
which our knowledge of dust optical properties has been derived. At longer
wavelengths, where the bulk of IR power is normally emitted, and is dominated
by grains big enough to be in thermal equilibrium with the radiation field,
the predictions are much more robust. Here, the largest source of uncertainty is
possibly the wavelength decline of the grain absorption coefficient.
The {\sl canonical} computations by \cite{draine84} used in this work yield
a power-law decline $\propto \lambda^{-\beta}$ with $\beta=2$ for $\lambda \gtrsim 40 \ \mu$m,
however several  laboratory measurements suggest a temperature dependence of $\beta$,
with $1.5 < \beta < 2.5$ \citep[see discussion and references in][]{jones14}.
Specifically, the possibility $\beta <2$ has sometimes been
welcomed,  {\rev since it provides some minor help} in the well known difficulty of galaxy formation
models in reproducing the high levels of bright sub-mm number counts
\cite[e.g.][]{baugh05}.

A detailed study of the dependence of IR  properties of simulated clusters
on dust properties is beyond the scope of this exploratory study.
However, in view of the discussion  of Section \ref{subsec:planck}, which
exploits the expected luminosity in the spectral range $\sim 100 - 400 \ \mu$m rest frame,
we checked that by adopting the extreme value $\beta=1.5$ the SED is
essentially unaffected below $100\ \mu$m. Above this wavelength
the expected luminosity is progressively enhanced, typically by about 30-40\%
at $400 \ \mu$m. This moderate increase would not affect any of our conclusions.

\begin{figure}
\hspace{-1cm}
\includegraphics[width=8.5cm, height=7.5cm]{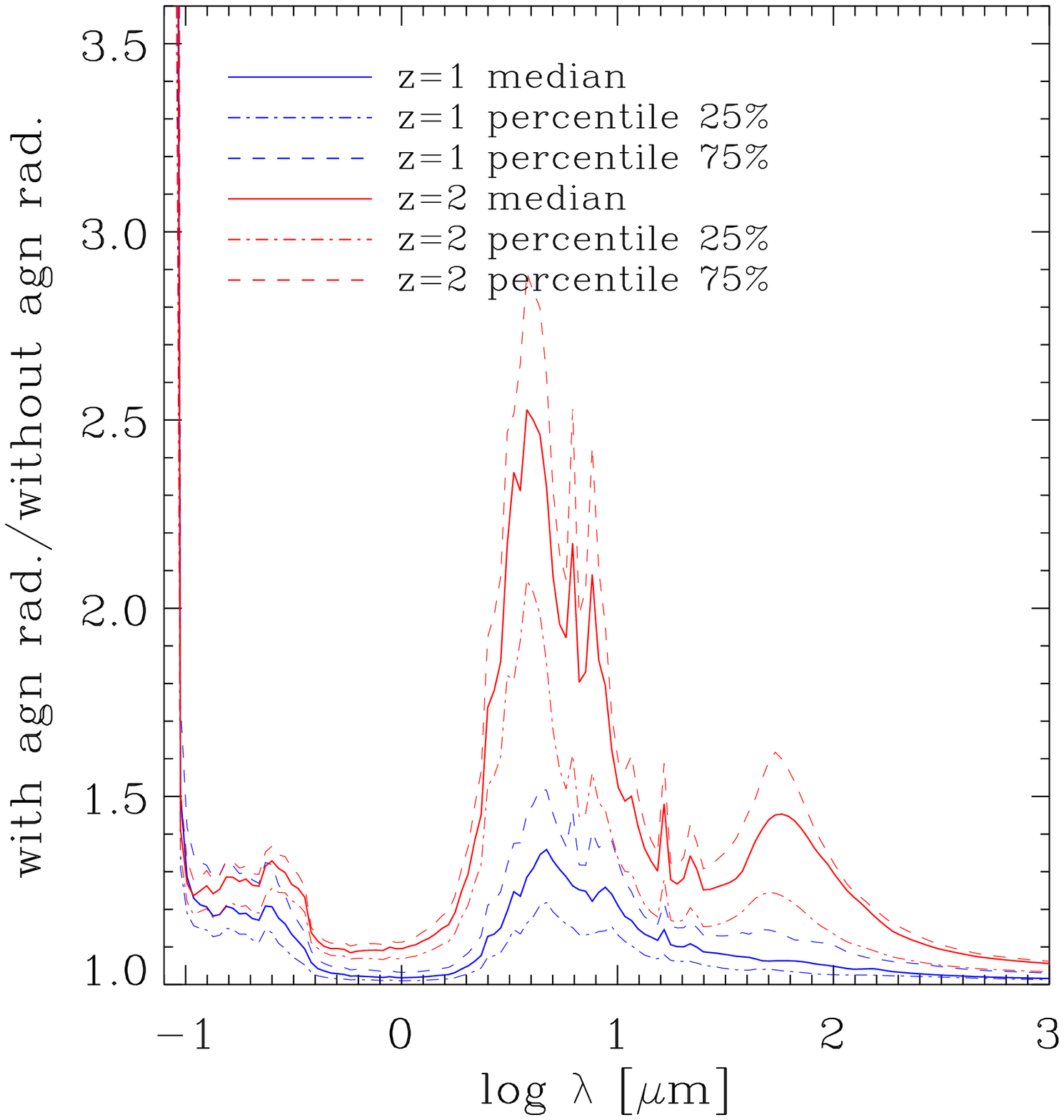}
\caption{Ratio of the predicted fluxes including or excluding the radiative
effect of AGNs, as a function of wavelength $\lambda$ (rest frame),
for our standard choice of \gtd\ parameters. Above $\sim 100 \ \mu$m
the difference is in most cases $\lesssim 25 \%$, decreasing
with increasing $\lambda$.}
\label{fig:with_over_without}
\end{figure}

%\clearpage

\begin{figure*}
\hspace{-1cm}
\includegraphics[width=8.5cm, height=7.5cm]{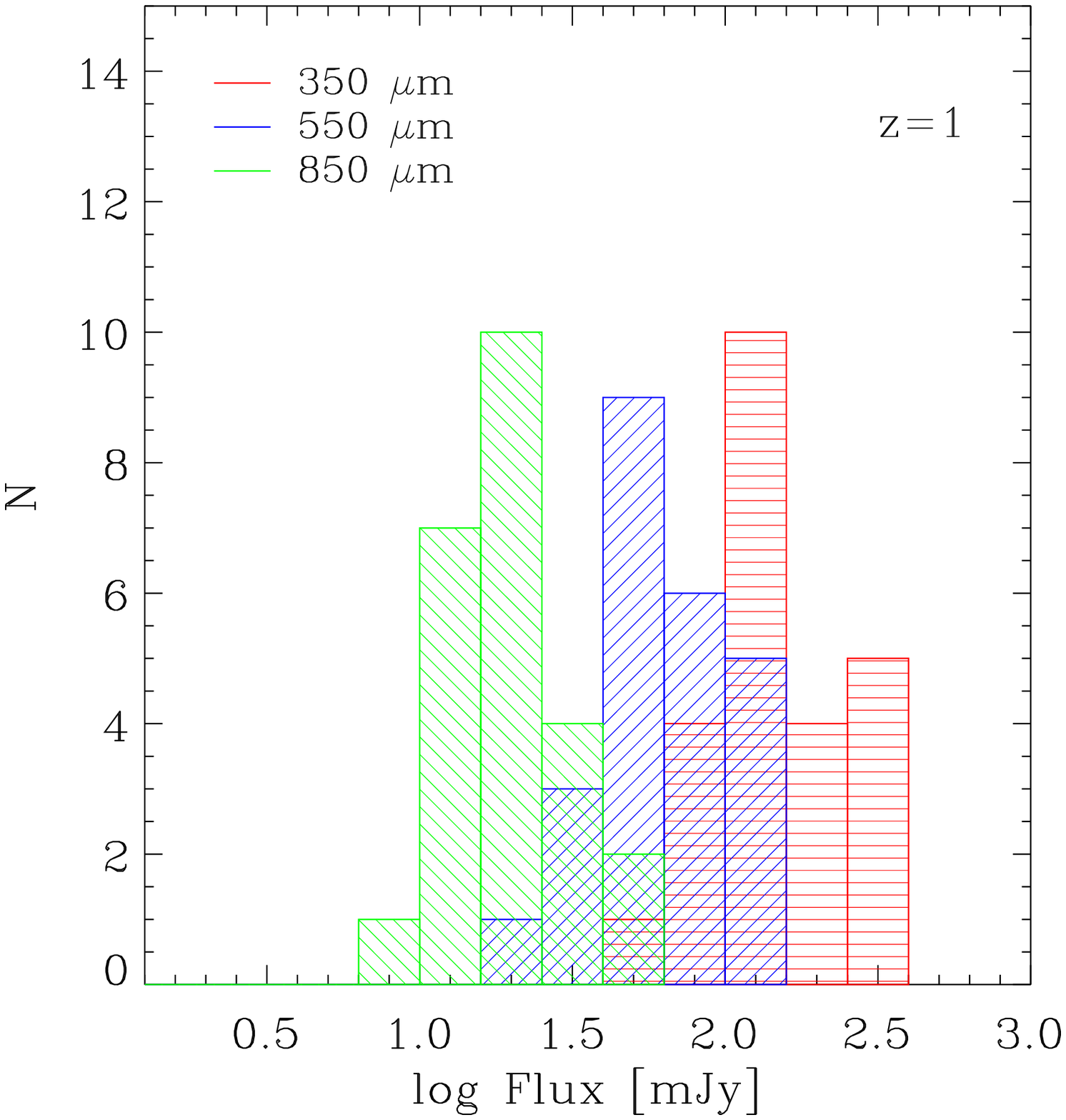}
\includegraphics[width=8.5cm, height=7.5cm]{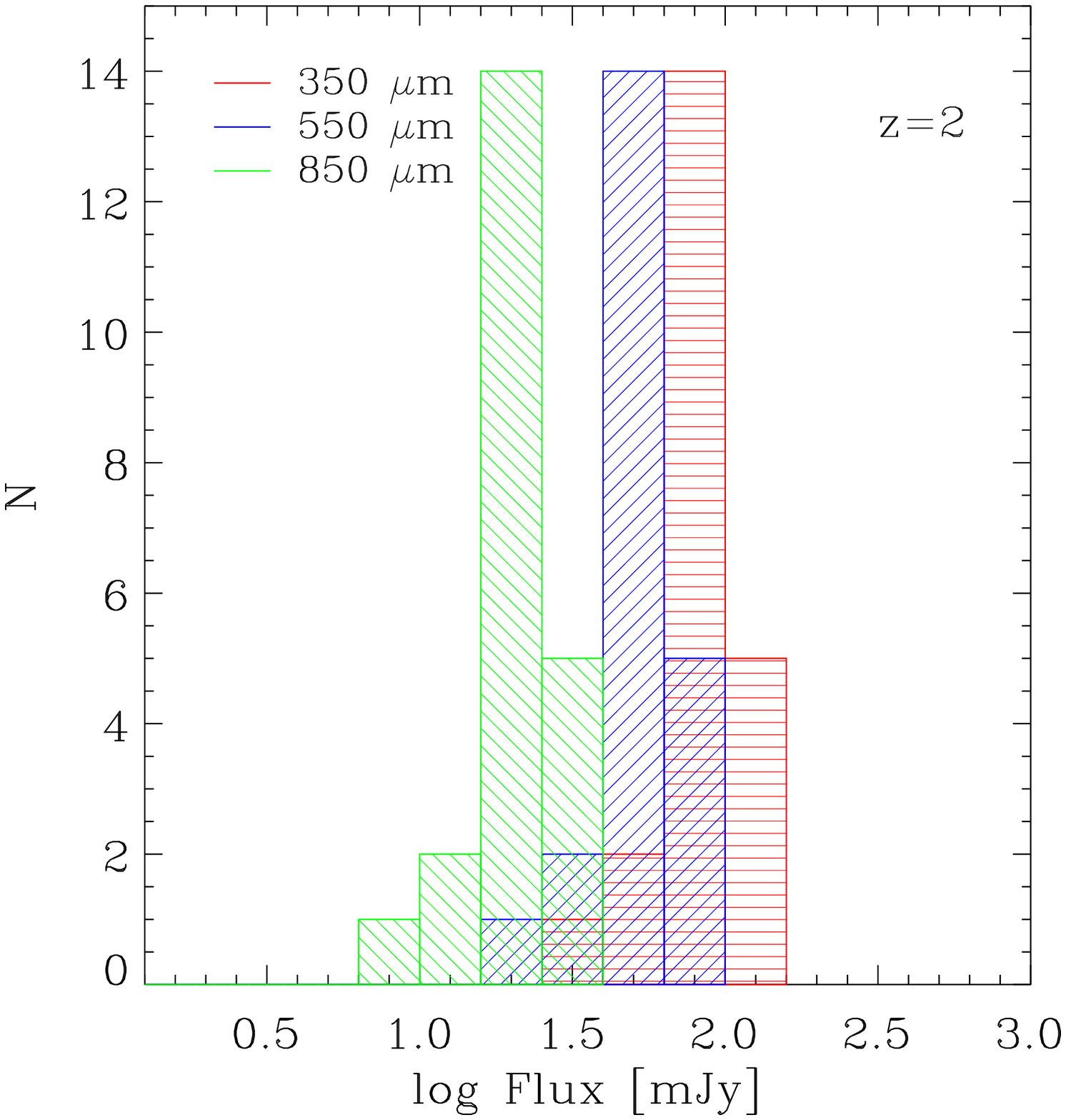}
\caption{Distributions of the predicted fluxes from our sample of simulated
clusters, within the Planck beam ($\sim$ 5 arcmin FWHM) in the 350, 550 and 850
$\mu$m HFI bands (857, 545 and 353 GHz respectively). The four candidate proto-clusters
selected by Clements et al.\ 2014
have photometric redshift of 0.76, 1.04, 2.05, and 2.26
and 350 $\mu$m HFI fluxes of 1100, 810, 1250 and 1240 mJy, i.e.\ close to the right
bound of these plots. However, these fluxes are believed to be
overestimated by a factor 2 to 3 due to selection bias. See text for details and discussion.}
\label{fig:his_flu_planck}
\end{figure*}

\begin{figure}
\hspace{-1cm}
\includegraphics[width=8.5cm, height=7.5cm]{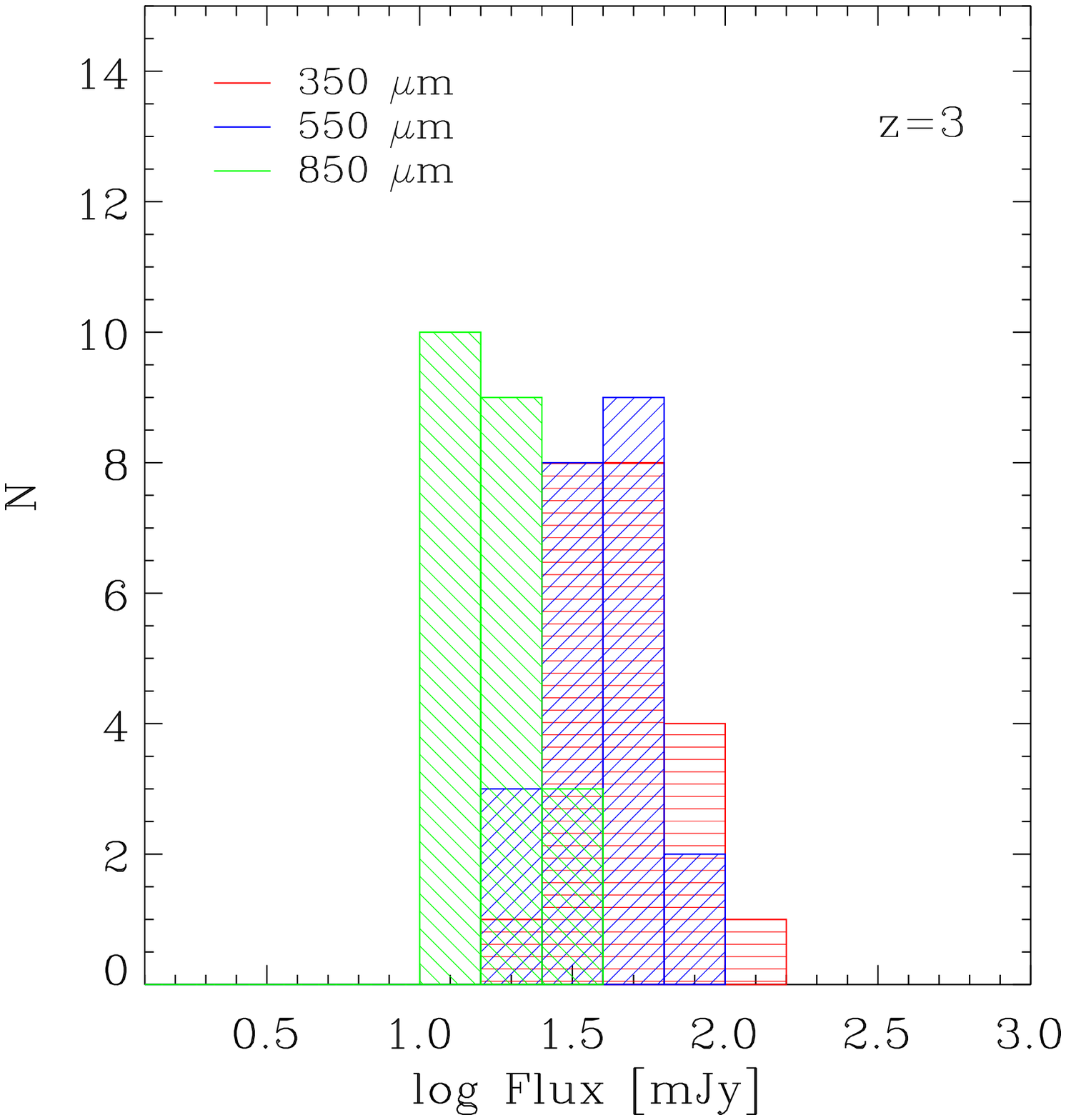}
\caption{Same as Figure \ref{fig:his_flu_planck},   but at z=3}
\label{fig:his_flu_planck_z3}
\end{figure}

\subsection{The radiative effect of SMBHs}
\label{subsec:eff_SMBHs} Fig. \ref{fig:cfrt_sed_radoff} and Fig.
\ref{fig:with_over_without} highlight the radiative contribution of the SMBHs
particles to the SEDs. The former one shows a comparison of the SEDs obtained
with or without their effect for a couple of clusters seen at z=1 and 2.
Also, the SEDs are computed for two very different AGN templates (shown in Figure \ref{fig:templates};
see Section \ref{sec:g3d}).
The latter figure displays the median ratio, the 25\% and the 75\% percentiles of the
predicted fluxes including or excluding the radiative effect of AGNs, for all
the sample. We note explicitly that in both cases we are considering the same simulations
in which the AGN feedback is included. Usually, in the spectral ranges
between 0.1 to 1 $\mu$m and 15 to 100 $\mu$m rest frame, we found that the AGN activity boosts the
integrated flux by a fraction ranging from $\sim 10 \%$ to $50 \%$ at z=2.
Moreover, the effect tends to decrease with decreasing redshift, being often
negligible at z=1. However, in the far UV at $\lambda \lesssim 0.1 \, \mu$m and
in the near IR
at $2 \, \mu\mbox{m}\lesssim \lambda \lesssim 15 \, \mu\mbox{m}$ its contribution
becomes more significant. In the former regime, this is due to the fact that
QSOs are believed to emit a significantly harder spectrum than the stars
below the Lyman limit, but the detailed result is strongly dependent on the
adopted AGN template (this is the main reason why this range is not shown in
Fig. \ref{fig:with_over_without}). In the latter spectral regime, the reason
is that only very concentrated and extremely luminous sources may produce an
interstellar radiation field so intense to heat dust grains at $T \sim 1000$
K, which is required to get thermal emission peaking at $\lambda\ \mbox{a
few } \mu$m. As a result, at $\lambda \sim 3\ \mu$m, the AGNs often boost the
predicted flux by a factor of a few. On the other hand, at $\lambda\ \gtrsim
100\ \mu$m rest frame, the radiative effect of SMBH particles amounts to less than 25\%,
with our chosen (as well as any reasonable) AGN SED template.

\subsection{Sub-mm properties of the clusters}
\label{subsec:planck}

In the previous sections we have seen that in the far-IR $\lambda
\gtrsim 100 \, \mu$m the uncertainties introduced by the sub-resolution
modelling of the radiative transfer in the MCs are negligible. The
dust emission is mostly powered by star formation with a tiny contribution
from AGN activity. These are interesting points in view of the efficiency of
far-IR/sub-mm surveys in detecting high-z objects in a violent, dust
obscured, star forming phase \citep[e.g.][and references therein]{coppin06}.
Indeed, it has been proposed  to take advantage of
large area surveys performed in this spectral region to uncover pristine
evolutionary phases of cluster regions, undergoing simultaneous
star-bursts \citep[e.g.][]{negrello05}.

In this vein, \cite{clements14} have recently reported the potential
detection of four clusters of dusty, star forming galaxies at photometric
redshift 0.76, 1.04, 2.05 and 2.26 by examining the Herschel-SPIRE images of
Planck Early Release Compact Source Catalog {\rev(ERCSC; Planck Collaboration
2011)} sources over an area of about 90 sq. degrees. With our panchromatic
computation of the expected SED of the most massive cluster regions simulated
in a large cosmological box, it is interesting to check to what extent these
detections can be explained by our simulations.  In this respect, it
is worth pointing out that the comoving volume encompassed by the 90 sq. deg.
area over the $z=0.76-2.3$ redshift range is of about $0.6 \, h^{-1}$ Gpc$^3$
for the adopted cosmology, thus smaller than the comoving volume of the
parent simulation.

\begin{figure*}
\hspace{-1cm}
\includegraphics[width=8.5cm, height=7.5cm]{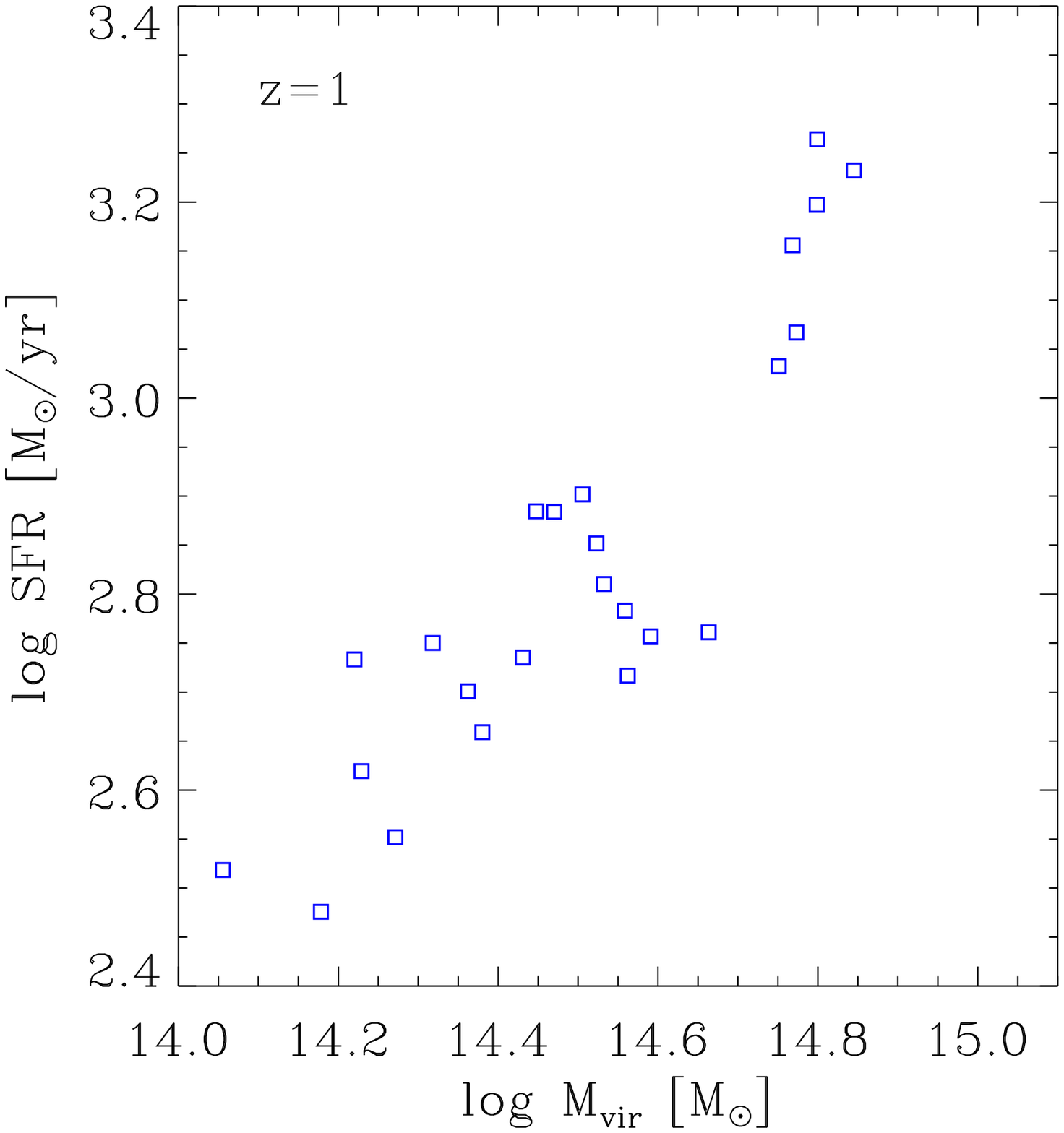}
\includegraphics[width=8.5cm, height=7.5cm]{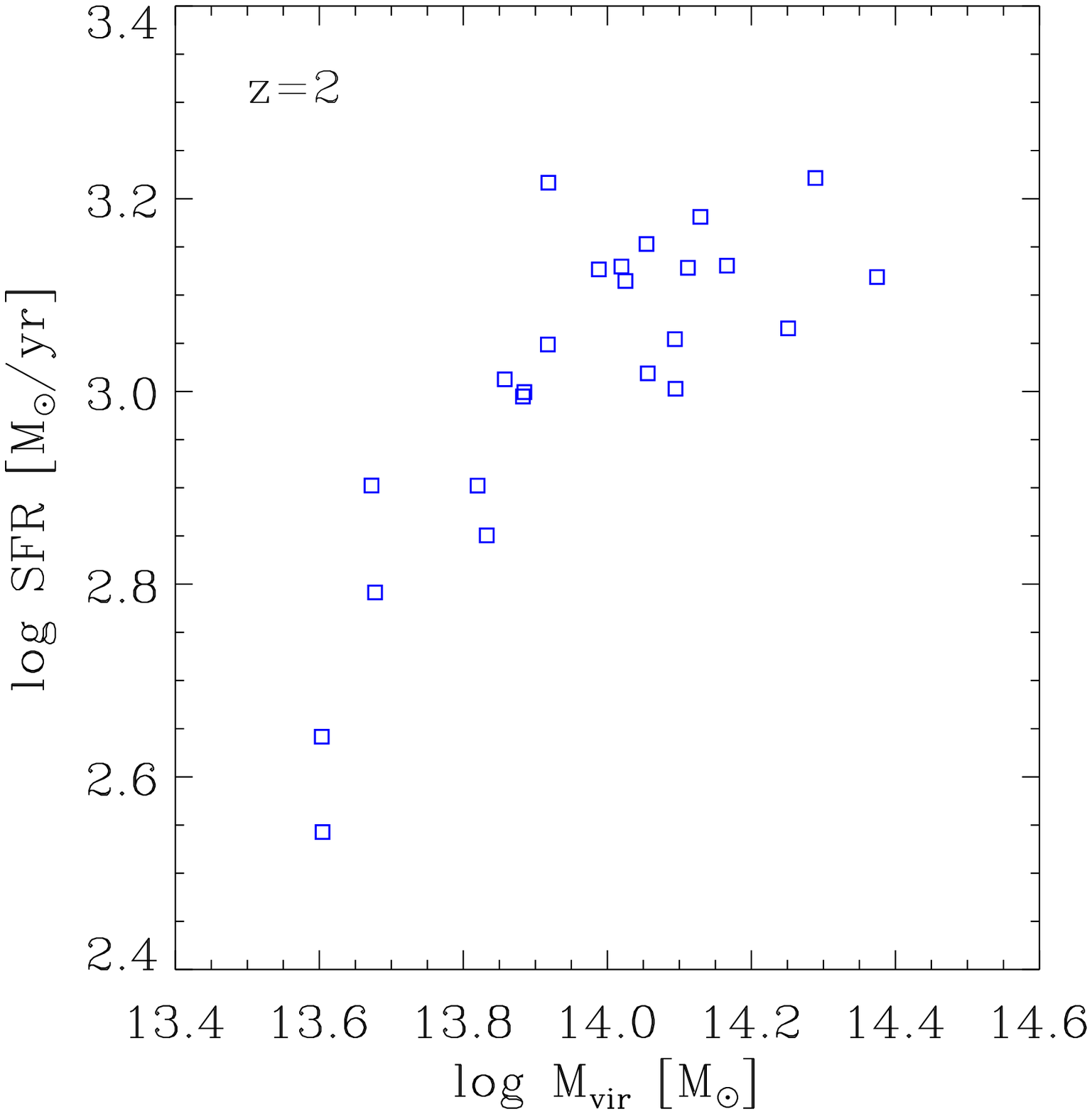}
\caption{SFR as a function of virial mass for the sample
of simulated clusters, at z=1 (left) and z=2 (right). The SFR
is calculated including all gas particles within a box of 2000 kpc,
close to the Planck HFI beam at both redshift.}
\label{fig:mass_sfr}
\end{figure*}

%{\rev For this purpose,
%it may be worth to point out that the parent box 1 Gpc $h^{-1}$, used to
%select the 24 most massive clusters formed at z=0, is big enough at any
%interesting redshift to cover the area examined by \cite{clements14}, since
%it corresponds to $\sim$ 560, 220 and 110 sq. degrees at $z=1,2,4$
%respectively.}

The measured Planck fluxes of the four cluster are all of the order of $\sim
10^3$ mJy at 350 $\mu$m (857 Ghz).  However, as mentioned by the authors
themselves, it must be taken into account that Planck fluxes fainter than 1.3
Jy are {\sl flux-boosted} due to the well known selection bias of faint
sources sitting on top of positive noise \citep{herranz13}. Indeed they found
that the sum of the fluxes of the Herschel sources in the Planck beam are
typically lower by a factor 2-3. Histograms of the predicted fluxes in the
Planck bands for our sample of mock clusters are shown in Figs.
\ref{fig:his_flu_planck} and \ref{fig:his_flu_planck_z3}. The median expected
flux is of about 131 and 85 mJy at z=1 and 2 respectively, but there are
cases reaching $\sim 300$ and $100$ mJy respectively. Thus, we can conclude
that the expected fluxes could marginally explain those reported by Clements
et al.\ for the two sources $z \sim 1$, assuming a  somewhat generous flux
boosting. On the contrary, they fail to do so  by a significative factor
$\gtrsim 3-4$ for the other two sources at $z \sim 2$. Indeed, while
\cite{clements14}, by considering the sum of the fluxes of Herschel sources
within the Planck beam, estimate for their two $z \simeq 2$ clumps SFRs of
{\rev at least}\footnote{For consistency, we decreased these numbers by a
factor 1.7 with respect to those reported by \cite{clements14}, since they
uses a calibration of the relationship between total IR luminosity and SFR
based on a Salpeter IMF, while our simulations adopt a more top heavy
Chabrier IMF (see also footnote \ref{note:imf}). However, Clements et al.\
claims that their SFRs are likely to be underestimated by a factor between
1.3 and 2. Note also that, had we adopted a Salpeter IMF in the simulations,
both the SFR level would be slightly reduced due to lower recycled fraction,
and the expected flux for a given SFR would be decreased, worsening both
discrepancies. } $\sim 2.9 \times 10^3$ and $7 \times 10^3$ \msunyr. By
converse, none of our simulated cluster has a SFR exceeding 1700 \msunyr
within the same beam (1300 \msunyr within the virial radius), with  median
value of 1000 \msunyr (800 \msunyr within the virial radius), as measured
directly on the simulation output (see Fig.\ \ref{fig:mass_sfr}). It is also
interesting to point out that very recently \cite{dannerbauer14}, using
APEX-LABOCA observations at 870 $\mu$m, determined a SFR of $\gtrsim$ 6300
\msunyr \ within a region of the same size $\sim$ 2 Mpc, around the
protocluster region traced by the {\sl spiderweb} radiogalaxy at $z=2.16$,
similar to that estimated by Clements et al., for their clumps at $z\sim 2$.
We recall that, as discussed in Section \ref{subsec:eff_SMBHs}, the
contribution from AGN power to the flux in the spectral region covered by
Planck is minimal for our simulated clusters. On the other hand, for the
other two observed clumps, whose photometric redshifts are 0.76 and 1.04,
their SFR estimates, {\rev again corrected downward to our adopted Chabrier
IMF}, are $\gtrsim$ 350 and 950 \msunyr respectively. These figures are thus
not inconsistent with the SFRs in our sample at z=1, whose median is 570 and
maximum $1700$ \msunyr.

\begin{figure}
\hspace{-1cm}
\includegraphics[width=8.5cm, height=7.5cm]{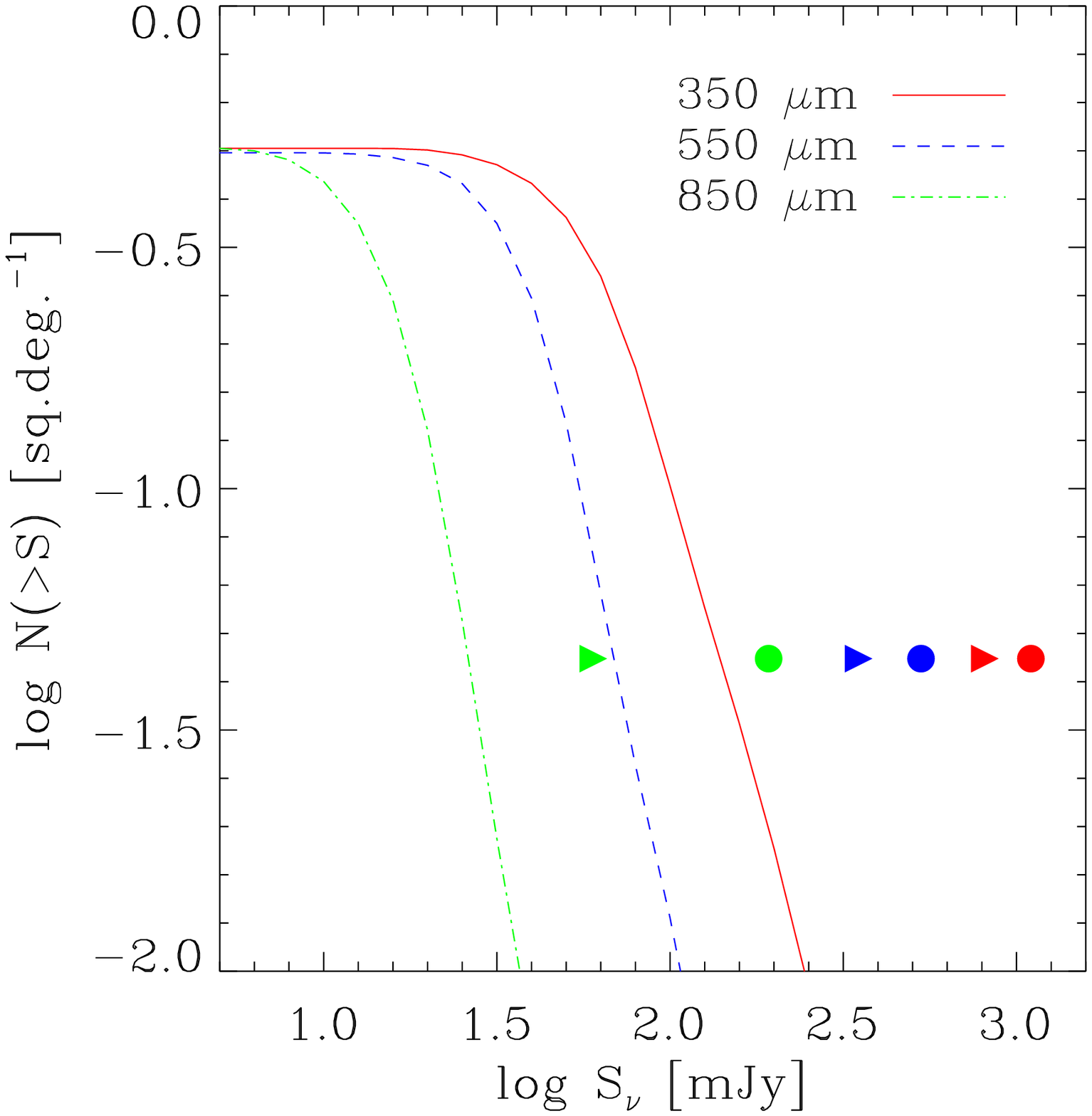}
\includegraphics[width=8.5cm, height=7.5cm]{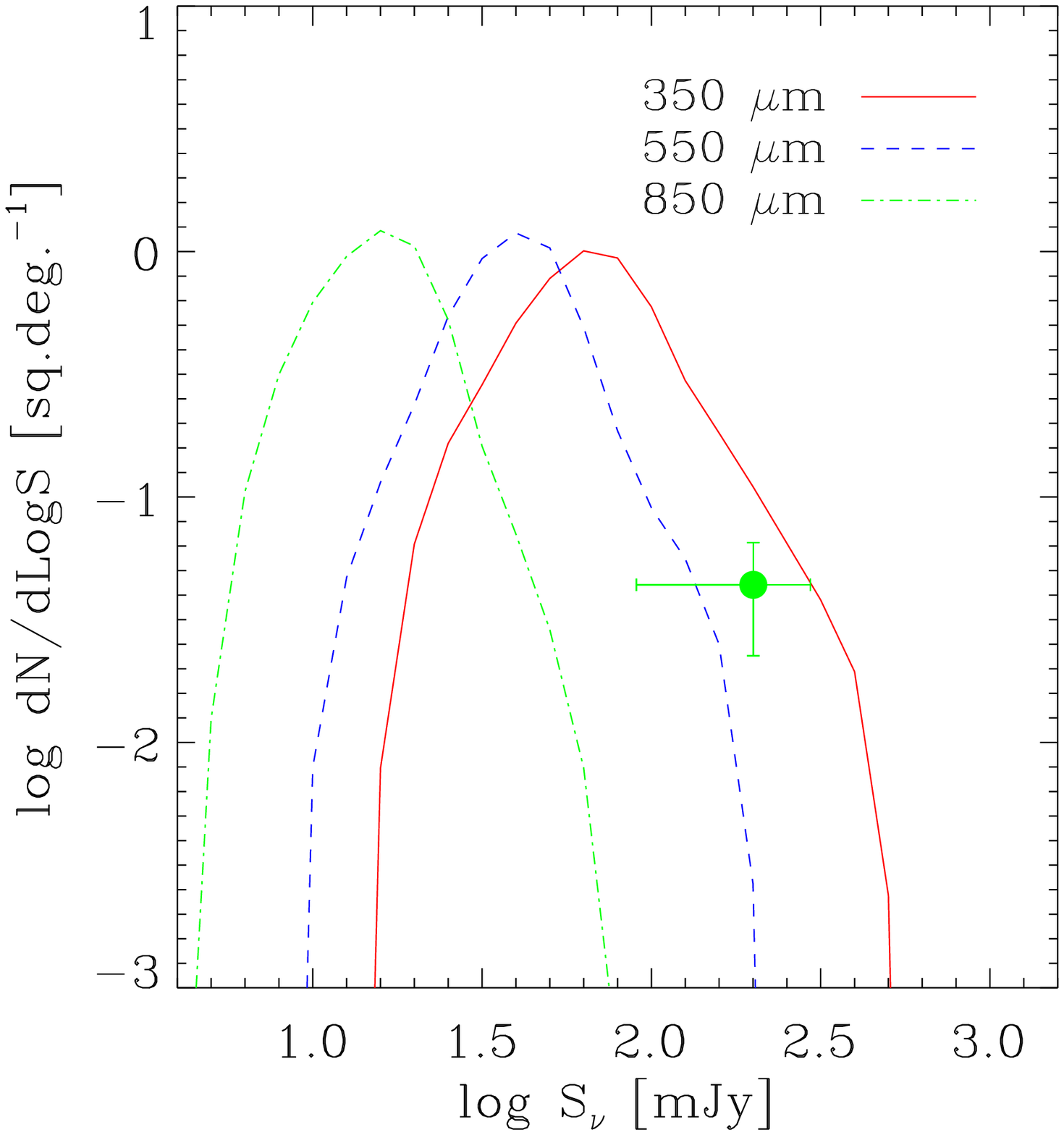}
\caption{Expected contribution to the cumulative and differential number counts
in the Planck HFI bands from the simulated clusters
considered in this work, namely all clusters whose final virial mass
$\gtrsim  1 \times 10^{15} \hmass$, treated as unresolved sources.
In the upper panel the colored circles (triangles) mark the position corresponding to the
average (minumum) fluxes and the number density
suggested by the four Planck clumps detected by Clements et al.\ 2014 over an area of 90 square degrees.
In the lower panel the circle with error bars is the estimate given by the these authors at 850 $\mu$m.}
\label{fig:counts}
\end{figure}

\begin{figure*}
\hspace{-1cm}
\includegraphics[width=9.5cm]{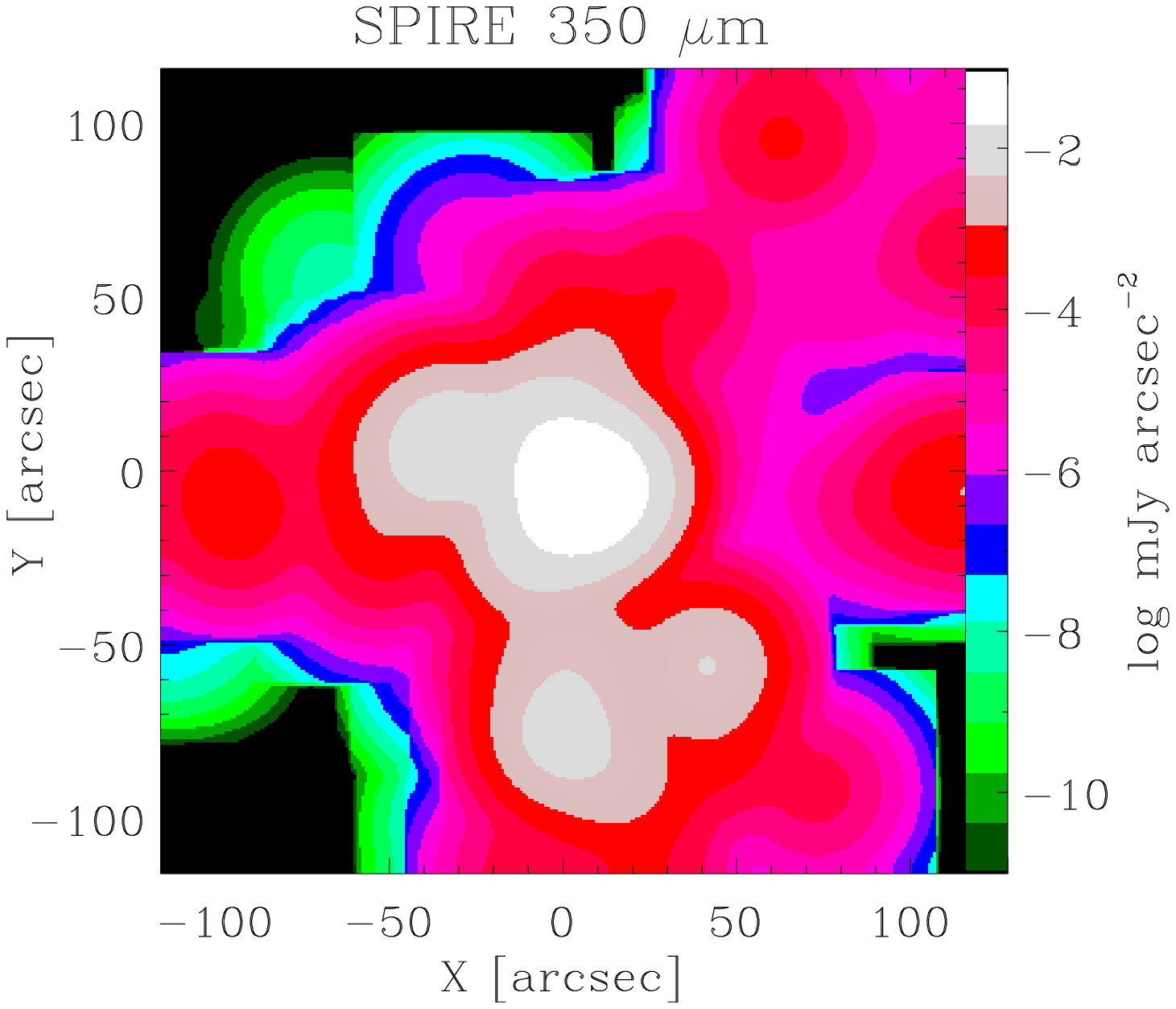}
\hspace{-1.2cm}
\includegraphics[width=9.5cm]{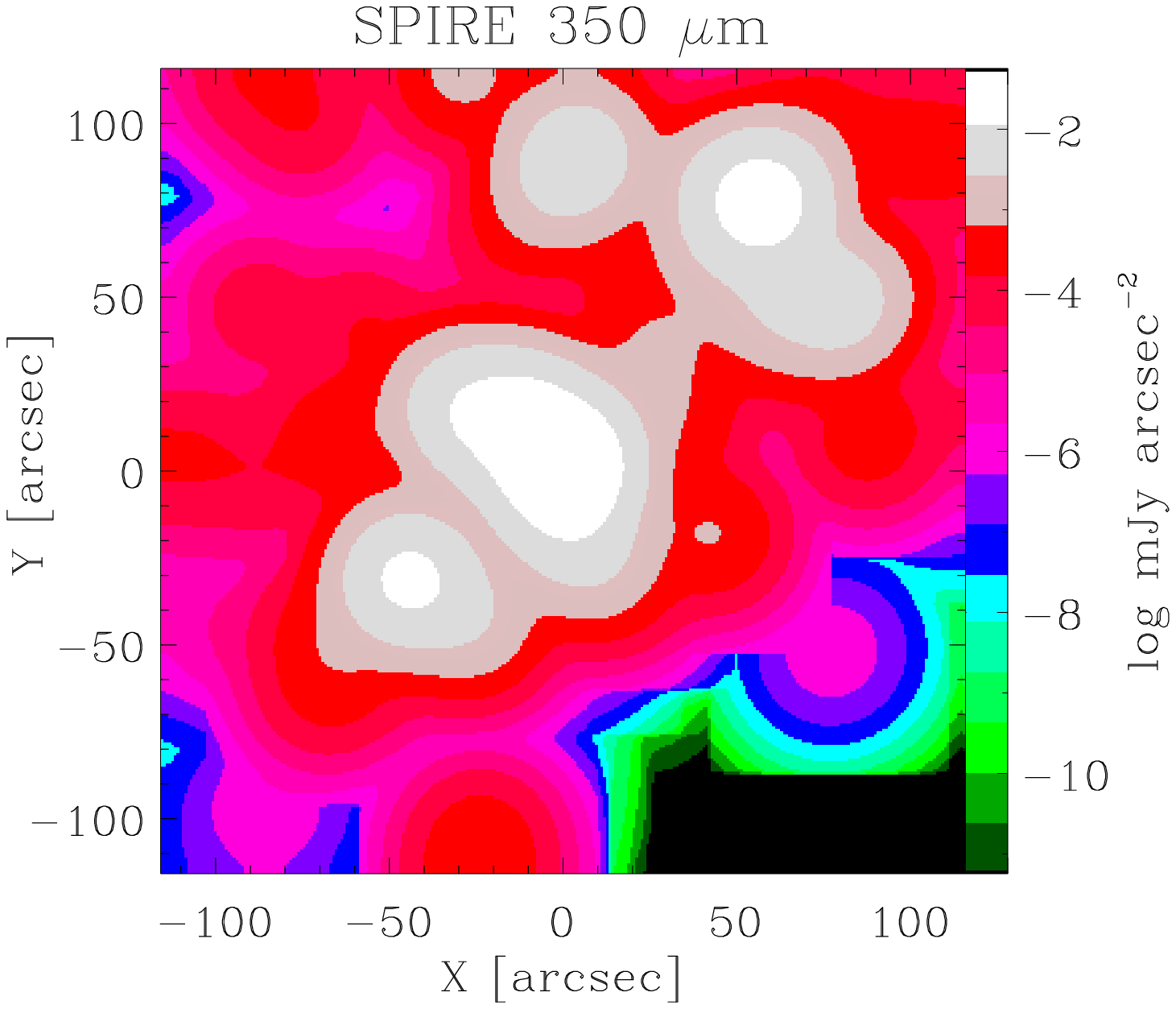}

\vspace{0.3cm}
\hspace{-1cm}
\includegraphics[width=9.5cm]{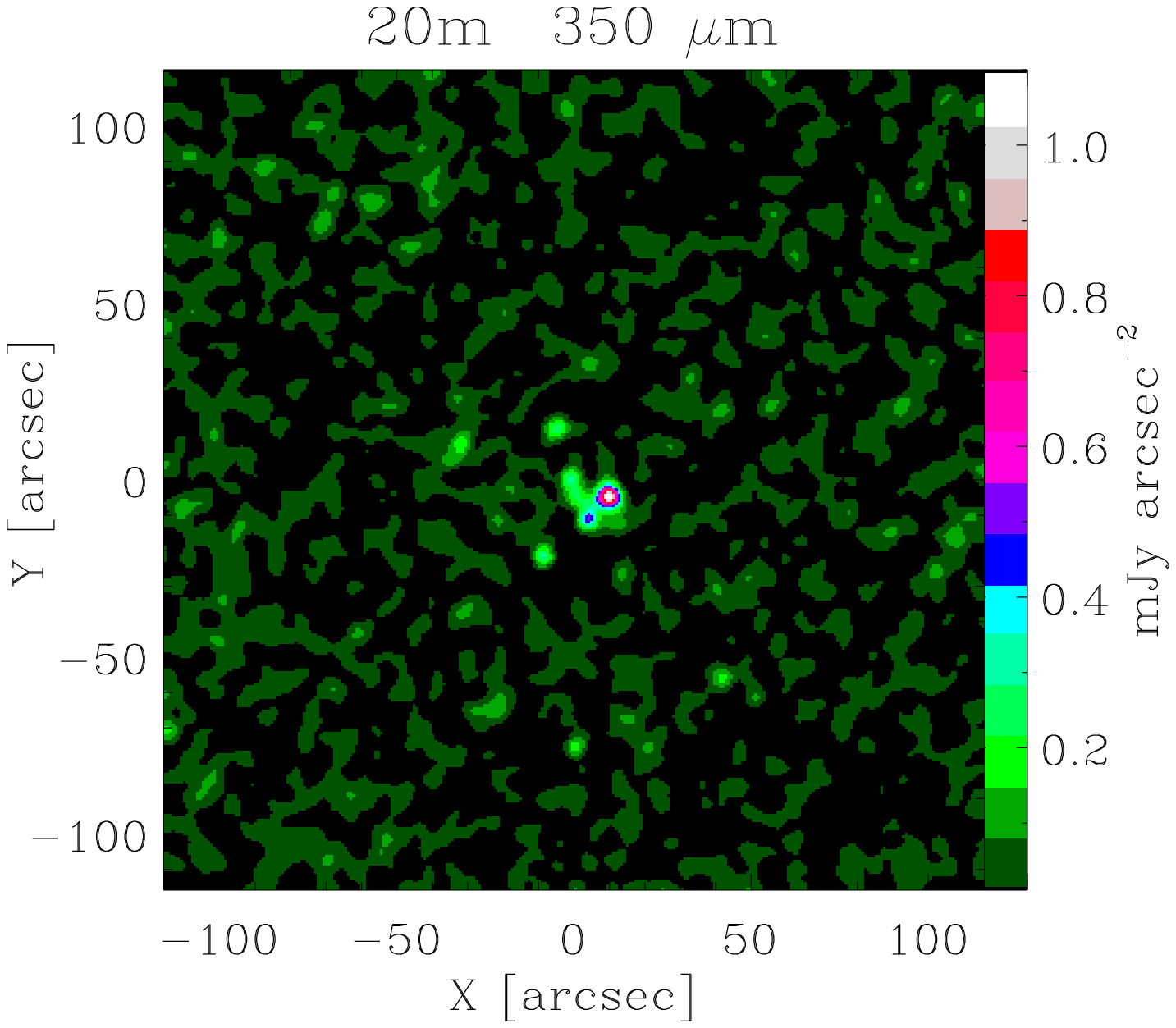}
\hspace{-1.2cm}
\includegraphics[width=9.5cm]{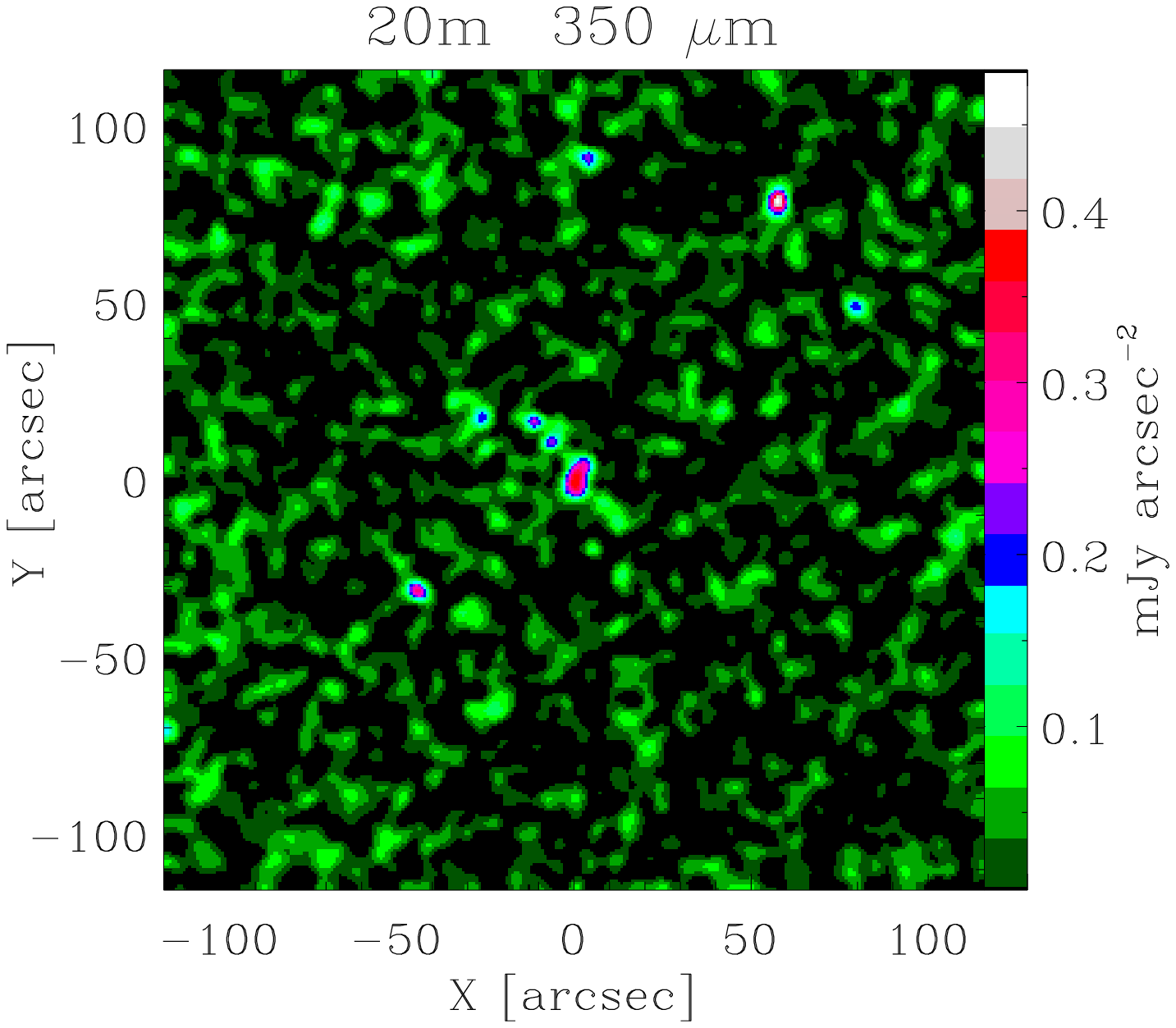}
\vspace{0.5cm}
\caption{350 $\mu$m (observed) \gtd\
images for the two cluster regions (left and right columns) at z=2 having the highest
SFR (at that redshift) in our sample, namely $\sim$ 1600 \msunyr\ within the
box. The box physical size is 2000 kpc, close to the Planck HFI beam at z=2.
The images in the top row have been convolved with a gaussian PSF of 25 arcsec FWHM,
corresponding to the Herschel telescope diffraction limit.
We expect at most 2-3 individual sources with a flux in the beam above
the Herschel confusion limit at this wavelength ($\sim 6$ mJy).
In the bottom row, the images include a gaussian noise of 0.1 mJy arcsec$^{-2}$
$1\sigma$, and are convolved with a PSF of 3.1 arcsec FWHM.
This resolution can be achieved by a 20 m single dish telescope \citep[see][]{sauvage14}.
}
\label{fig:map_z2_convol}
\end{figure*}

%\caption{350 $\mu$m (observed) \gtd\
%images for the two cluster regions (left and right columns) at z=2 having the highest
%SFR (at that redshift) in our sample, namely $\sim$ 1600 \msunyr\ within the
%box. The box physical size is 2000 kpc, close to the Planck HFI beam at z=2.
%The numbers on the horizontal and vertical left axes marks arcsec.
%The images in the top row have been convolved with a gaussian PSF of 25 arcsec FWHM,
%corresponding to the Herschel telescope diffraction limit.
%Here, the color coded flux units are logarithm of mJy arcsec$^{-2}$.
%We expect at most 2-3 individual sources with a flux in the beam above
%the Herschel confusion limit at this wavelength ($\sim 6$ mJy).
%In the bottom row, the images are instead presented in linear flux units of
%mJy arcsec$^{-2}$, with the addition of a gaussian noise of 0.1 mJy arcsec$^{-2}$
%$1\sigma$, and  convolved with a PSF of 3.1 arcsec FWHM.
%This resolution can be achieved by a 20 m single dish telescope \citep[see][]{sauvage14}.
%}

Fig.\ \ref{fig:counts} shows the contribution expected from our simulated
clusters to the cumulative number counts, as unresolved sources, in the
Planck HFI bands. They have been computed by integrating in redshift
the luminosity functions evaluated at several redshifts between 0.5 and 3. As
expected from the previous discussion, they do not show up at all at flux
levels of the order of 1 Jy and in order to find a few clusters over an area
of $\sim 90$ square degrees, as reported by \cite{clements14}, it would be
necessary, according to our computations, to reach much lower sensitivities,
$\sim 100$ mJy at 350 $\mu$m. This is far too low even for the typical
sensitivity $\sim 0.7$ Jy of the final Planck catalogues of compact sources
\citep{planck13}. {\rev It may be worth pointing out that \cite{negrello05},
exploiting the simple semi-analytic model (SAM) by \cite{granato04}
for the co-evolution of spheroids and QSO, which
provided a good match to the sub-mm number counts available at that time,
estimated, for an instrument with the resolution of Planck/HFI,
surface densities of 850 $\mu$m sources in good keeping with the findings by
\cite{clements14}\footnote{See figure 14 in \cite{clements14}, bearing in mind that they dubbed
{\it numerical simulation} a Monte Carlo realization of sub-mm sky,
again based on the \cite{granato04} SAM.}. Furthermore, from the observational point of view, these finding
seems to be strengthened by a more recent combined analysis of much larger portions of the sky
observed by Planck and Herschel \citep{planck15}. In this case, hundreds of over-densities
of high-z galaxies have been identified, which may be interpreted as proto-clusters characterized
by typical global SFR of the order of several thousands \msunyr.}

{\rev Figure \ref{fig:mass_sfr} highlights that our sample of simulated cluster regions
is characterized by a strong correlation between the instantaneous virial mass and the SFR,
computed within a box matching the Planck-HFI beam, both at z=1 and 2. We find a similar well
defined correlation between the SFR at any interesting z and the {\it final} virial mass,
on which the selection of the sample is based (Section \ref{sec:clusel}). As a consequence,
it seems very unlikely that, extending our study to lower mass clusters, we could
find systems featuring high z SF activity as intense as that hinted by the
recent observations and never attained by our objects.
}

In Figure \ref{fig:map_z2_convol} we show the predicted z=2 images in the
Herschel-SPIRE 350 $\mu$m band  for two cluster regions. These are the two
regions characterized by the highest SFR in our sample at that redshift,
$\sim$ 1600 \msunyr\ within the box, and have been convolved with a gaussian
PSF of 25 arcsec FWHM, corresponding to the telescope diffraction limit. As
can be judged by these maps, the brightest simulated clusters are expected
to produce at most 2-3 individual sources with a flux in the beam slightly above
the Herschel confusion limit at this wavelength ($\sim 6$ mJy) \citep{nguyen10}.

In conclusion, it would be very difficult, if not impossible, to uncover and
study structures similar to those predicted by our simulations with present day FIR
and sub-mm space facilities.

As we mentioned before, we have also simulation runs with AGN feedback
switched off. In this case the predicted SFRs, and correspondingly the
expected fluxes in the Herschel bands, are typically higher by a factor $\sim
2$ and 1.5 at redshift 1 and 2 respectively. These numbers would be in better
agreement with the figures quoted by \cite{clements14}, albeit still too low
at $z\sim 2$. Moreover, it is well known that without some treatment of this
astrophysical process, cosmological simulations over-predict the amount of
baryons converted into stars, particularly in massive systems, by about one
order of magnitude \citep[see for instance][and references
therein]{ragone13}.

{\rev One possibility to increase the predicted fluxes would be to adopt a
more top heavy IMF. However, in order to get an increase by a factor
of a few, as required in particular by the finding of \cite{clements14},
the IMF should be quite extreme, similar to the flat IMF postulated
by the Durham SAM (during bursts) to reproduce the sub-mm number counts \citep{baugh05}.
We have verified that in this case the SFR would be only slightly modified,
because two competing effects almost cancel each other: more gas recycling and faster
chemical enrichment on one side, and stronger SNae feedback on the
other side. In the deepest developing potential wells
yielding to the formation of the most massive clusters, we found,
according to our prescriptions, some prevalence of the former effect, causing
an increase of  SFR by no more than 10-20 \%. However, adopting such an extreme IMF,
the predicted flux for a given SFR at $\lambda> 100 \mu$m rest frame can increase
by a factor of a few, the exact value depending on the star formation history.
There are some hints of IMFs more top heavy than Chabrier
in violently star-forming environments (but also for the opposite), but
nothing approaching such an extreme IMF
has ever been observed. Moreover, later studies have shown that this
quite ad hoc assumption leads to a few remarkable problems in the context
of the Durham SAM \cite[for a discussion see][]{casey14}.
It seems therefore unlikely that the difficulty highlighted in this section
can be entirely solved by the adoption of a different IMF in our simulations.
}

\section{Summary and conclusions}
\label{sec:conclu}

In this work we have post-processed a sample of cosmological zoom-in
simulations following the formation of the 24 most massive galaxy clusters
selected from a parent simulation of box size of $1 h^{-1} Gpc$. The final
virial mass of the clusters ranges from $\simeq$ 1 to $3 \times 10^{15}
\hmass$. The post processing consists of performing radiative transfer
computations with the \gtd\ code \citep{g3d} including dust reprocessing, in
order to predict the IR properties of the forming clusters during the most
active star forming phases. We have implemented in \gtd\ a treatment of the
radiative contribution due to AGN activity, consistent with the prescriptions
adopted in the simulation for the AGN feedback. The latter is widely
recognized as a key ingredient to limit the overproduction of stars in
massive halos. The expected contribution to the IR emission from accretion
power could be significant at $\lambda \lesssim 100\ \mu$m, but minor or
negligible at $\lambda \gtrsim 100\ \mu$m. However,
going to shorter and shorter wavelengths, the
exact budget  becomes progressively dependent on the adopted \gtd\ assumptions.

We have demonstrated that during the early phases of assembly of massive
galaxy clusters our simulations do not reach far IR luminosities high
enough, by a factor at least of a few, to account for the reported discovery of
four high z, massively star forming clusters by \cite{clements14}, over an
area of about 90 sq. degrees in the Planck satellite survey.  Since we have
shown that this conclusion is very robust with respect to any reasonable
variation of the assumptions required to perform the
dust emission computations and that the possible contribution to the overall
emission from AGN is small in this spectral regime, the problem directly
translates to {\sl insufficient peaks} of star formation activity in the
simulations at early epochs. This problem becomes more puzzling taking into
account that the same simulations {\sl over-predict} the final stellar mass
in BCGs hosted by massive clusters at z=0, by a factor of a few
\citep{ragone13}.

%A simple recalibration of the sub-grid prescription,
%or an increase of numerical resolution \cite[e.g.][]{borgani06}, in such a way to increase the high-z
%star formation activity would worsen this latter tension.

High-redshift star formation rates can be increased in numerical simulations.
Improving the resolution would already boost the SFR to some extent
\cite[e.g.][]{borgani06}, but still not enough to eliminate the tension between
observations and the models. Also, it would be possible to re-calibrate the
sub-grid prescriptions for the baryon physics. However, it appears very
difficult to enhance the SFR at high redshifts without also increasing the
final mass of simulated BCGs, thus worsening the above mentioned
disagreement. {\rev Indeed there have been already reports of
a paucity of strong starburts in recent cosmological simulations,
once their sub resolution physics
is calibrated to reproduce other constraints \cite[e.g.][]{sparre14}.
These violent star formation events
would be required to explain the statistical properties of the general
population of sub-mm selected galaxies (SMGs).
We wish to address this issue in the near future, by applying the \gtd\
post processing described in this paper to a forthcoming hydrodynamical
simulation of an entire (rather than just zooms of the massive cluster regions)
cosmological box, including sub-resolution physics. This would provide
new clues on the ongoing debate on the use of SMGs to trace the assembly of
massive structures at high-z \citep[e.g.][and references therein]{miller15}.}

Moreover, recent observations suggest a picture according to which the $z
\gtrsim 1.5$ population of galaxy (proto)clusters contains examples with both
extreme \citep[e.g.][]{tran10,clements14,dannerbauer14,santos14,santos14b} as well as
very low star formation activity \citep[e.g.][]{tanaka13,kubo14}, while in
our simulated clusters these two opposite situations are clearly
under-represented. It seems that the bulk of star formation in the
progenitors of real massive galaxy clusters occurred at higher rates, but
lasted less than in our simulations.

These opposite tensions may indicate that the prescriptions adopted to
describe the sub-resolution processes should be improved to better capture
the relevant physics. Possibly, if the situation hinted by the growing data
set on (proto)clusters at $z \gtrsim 1$ is confirmed, the interplay between
the feedback schemes and the star formation model should yield overdense
regions characterized by both higher and lower star formation levels than
attained in current simulations, while avoiding an overproduction of stars
over the assembly history of the most massive cluster galaxies. It is
conceivable that an early phase of positive feedback during the development
of AGN activity \citep[e.g.][]{silk13} and/or a longer delay between the
most violent episodes of star formation and the onset of efficient AGN
quenching \citep[e.g.][]{granato04} could be part of a more realistic
modeling.

Finally, we notice that since our prescriptions for the baryonic sub-grid
physics, apart minor variations, are quite standard in present day
cosmological simulations, we expect that our findings would apply to most
of them in the same halo mass regime.

\section*{Acknowledgements}
We thank the referee Dave Clements for his constructive comments, which substantially helped
to improve the quality of the paper. We acknowledge useful exchange of ideas
and information with Gianfranco
De Zotti, Herve Dole, Samuel Farrens, Chris Hayward, Paolo Tozzi and Joana Santos.
G.L.G.\ acknowledges warm hospitality by IATE-C\'ordoba during the development
of the present work. The authors thank Volker Springel for making available to us
the non-public version of the GADGET–3 code. Simulations have been carried
out at the CINECA supercomputing Centre in Bologna, with CPU time assigned
through ISCRA proposals and through an agreement with University of Trieste.
C.R.F.\ acknowledges founding from the  Consejo Nacional de
Investigaciones Cient\'{\i}ficas y T\'ecnicas de la Rep\'ublica Argentina
(CONICET), by the Secretar\'{\i}a de Ciencia y T\'ecnica de la Universidad
Nacional de C\'ordoba (SeCyT) and by the Fondo para la
Investigaci\'on Cient\'{\i}fica y Tecnol\'ogica (FonCyT).
This work has been supported by the PRIN-MIUR 201278X4FL “Evolution
of cosmic baryons” funded by the Italian Ministry of Research, by the
PRIN-INAF09 project “Towards an Italian Network for Computational Cosmology”,
and by the INDARK INFN grant, by the MICINN and MINECO (Spain) through the grants
AYA2009-12792-C03-03 and AYA2012-31101 from the PNAyA and by the
European Commission's Framework Programme 7, through the International
Research Staff Exchange Program LACEGAL. A.O.\ was financially
supported through a FPI contract from AYA2009-12792-C03-03.

%- commentare che abbiamo pure provato a "truccare" la emissivita' della polvere
%per alzare i flussi?

%- quanto bisognerebbe far salire AGN per farlo dominare a lunghezze d'onda Clements?

{}

%\clearpage

%\appendix
%\section{A}
%\label{AppA}

%In this Appendix...

\end{document}